\documentclass[12pt]{article}
\usepackage{amssymb}
\usepackage{epsfig}

\def\be{\begin{equation}}
\def\eeq{\end{equation}}
\def\bea{\begin{eqnarray}}
\def\eea{\end{eqnarray}}

\def\re#1{(\ref{#1})}

\begin{document}

\begin{titlepage}
\vskip 2 cm
\begin{center}
\huge{\bf{Kaluza-Klein towers in warped spaces with metric singularities}}
\end{center}

\begin{center} {\Large{Fernand Grard$^1$}},
\ \ {\Large{Jean Nuyts$^2$}}
\end{center}
\vskip 0.5 cm

\noindent{\bf Abstract}
\vskip 0.2 cm
{\small
\noindent
The version of the warp model that we
proposed to explain the mass scale hierarchy
has been extended by the introduction 
of one or more singularities in the metric. 
We restricted ourselves to a real massless scalar field 
supposed to propagate in a five-dimensional bulk 
with the extra dimension being compactified on a strip or on a circle. 
With the same emphasis on the hermiticity and 
commutativity properties of the Kakuza Klein 
operators, we have established all the allowed 
boundary conditions to be imposed 
on the fields. From them,
for given positions of the singularities, 
one can deduce either mass eigenvalues 
building up a Kaluza-Klein tower, or a tachyon, 
or a zero mass state. 
Assuming the Planck mass to be the high mass scale and 
by a choice, unique for all boundary conditions, of the major warp parameters, 
the low lying mass eigenvalues are of the order of the TeV,
in this way explaining the mass scale hierarchy.
In our model,
the physical masses are related to the Kaluza-Klein eigenvalues, 
depending on the location of the physical brane 
which is an arbitrary parameter of the model.
Illustrative numerical calculations are given 
to visualize the structure of Kaluza-Klein mass eigenvalue towers. 
Observation at high energy colliders like LHC of a mass tower 
with its characteristic
structure would be the fingerprint of the model. 
}

\vfill
\noindent
{\it $^1$  Fernand.Grard@umh.ac.be,  Physique G\'en\'erale et
Physique des Particules El\'ementaires,
Universit\'e de Mons-Hainaut, 20 Place du Parc, 7000 Mons, Belgium}
\vskip 0.2 cm
\noindent
{\it $^2$  Jean.Nuyts@umh.ac.be,
Physique Th\'eorique et Math\'ematique,
Universit\'e de Mons-Hainaut,
20 Place du Parc, 7000 Mons, Belgium}
\end{titlepage}

\pagestyle{myheadings}
\markboth{}{\today}

\section{Introduction \label{Intro}}

In our analysis 
of the procedure of 
generation of Kaluza-Klein masses for scalar fields in a 
five-dimensional flat space with its fifth dimension compactified
\cite{GN1},
we stressed that it is the momentum squared in the extra dimension, at the 
basis of the Kaluza-Klein reduction equations, which must be an hermitian operator and not 
the momentum itself. This resulted in the establishment of 
specific boundary 
conditions to be imposed on the fields of interest. 
Similar considerations have been applied to the case of spinors fields
in a five-dimensional flat space in \cite{GN3}.

Following the same line of thought
and inspired by the Randall Sundrum scenario \cite{RS2}, we developed a warp model 
\cite{GN2}
adopting their metric for a space
with a bulk negative cosmological constant 
in view of solving the mass hierarchy problem. 

This warp model
has been elaborated independently in a mathematically consistent and complete way,
up to dynamical considerations.
Restricting ourselves to a real massless scalar field supposed to propagate in the 
bulk, we postulated
that the fifth dimension is compactified on a strip (of 
length $2\pi R$) and, 
in this first version, that the metric has no 
singularities.
As in \cite{GN1}, after a careful study of the hermiticity and commutativity properties of the 
operators entering in the Kaluza-Klein reduction equations, 
we have enumerated 
all the allowed boundary 
conditions. From them, we have deduced the mass eigenvalues corresponding 
to the Kaluza-Klein towers and tachyon states.

The basic assumption in the model is that there is one mass scale only, the 
Plank mass. By an adequate choice, agreeing with this assumption, of the 
two major parameters of the model, namely the warp factor $k$ and $R$, 
it turns out that the low lying Kaluza-Klein 
mass eigenvalues 
can be made of the order of one TeV, 
solving in this way the mass scale hierarchy 
problem.  This result holds true for all boundary conditions at once
without fine tuning.

A specific aspect of our warped model is that the physical masses as 
observed in the TeV brane (the physical brane in 
which we live) can be deduced from the mass eigenvalues.
They 
depend on the location of that brane 
on the extra dimension axis. This location is an arbitrary parameter
of the model.

In this article, still considering
the 
case of a real massless scalar field propagating 
in the bulk with the fifth dimension 
being compactified on a strip, we extend our warp model by the consideration 
of one (see Sec.\re{General}) or more (see Sec.\re{anysing}) metric singularities.
The metric singularities (see Sec.\re{metricsec}) are located at some fixed points on the 
extra dimension axis where the metric components 
are continuous but not their derivatives.
Under some specific conditions, the strip can be closed into a circle 
(with an even number of singularities \re{circle}).

In the main part of the paper,
the model is extended  along the same lines as in our 
previous articles, first when there is a single singularity.
We start
from the Riemann equation (see Sec.\re{eqmotion})
which results
from a least action principle with the usual Lagrangian,
when the field variations are taken to be zero
at the boundaries of the domains.
In Sec.\re{herm},
we put the same emphasis on the hermiticity and commutativity properties
of the relevant Kaluza-Klein operators (see Sec.\re{KKreduction})
which originate from the Riemann equation and guarantee that the mass eigenvalues are real.
The field boundary conditions are established in Sec.\re{HBC}
as a generalisation of  the ones valid in the case of 
a metric without any singularity \re{GN2}.
These boundary conditions are seen to be compatible with those obtained 
from the least action principle with a more general
Lagrangian leading to the same Riemann equation 
under the hypothesis that the fields and their variations belong to the same Hilbert space
(See App.\re{least})
The solutions for the fields together with the related mass eigenvalue equations are 
formulated in Sec.\re{solutionform}. 
The physically important case of a zero mass eigenvalue is treated in Sec.\re{zeromassBC}. 

The results of Sec.\re{General} have been generalized 
in Sec.\re{anysing} for an arbitrary number $N$ of singularities.
The hermiticity properties and boundary conditions are discussed in Sec.\re{hermN}, the 
Riemann equation solutions and mass eigenvalues in Sec.\re{anysingeigen}.  
The closure of the strip into a circle is treated in Sec.\re{circle}.

In Sec.\re{physdis}, a few physical considerations are 
made in relation with the results presented in the two preceeding sections. A discussion
of the meaning 
of our boundary conditions is carried out in Sec.\re{physdis1}.
With the same assumption that the Planck mass is the only mass scale 
in the problem,  
the physical interpretation of the Kaluza-Klein mass eigenvalues is 
conducted in Sec.\re{Plscale}
in a completely analogous way as without any singularity. 
The only price to pay is that the choice 
of the major parameters $k$ and $R$ must 
depend on the location of the singularities 
in order to protect the mass hierarchy.
Moreover, what we considered as the specific aspect of our model, namely the relation 
between
the Kaluza-Klein mass eigenvalues and the physical masses, remains valid (Sec.\re{physicmass}).
As a consequence, 
they both depend on the
positions of the singularities. A few words are devoted to the probability densities along the 
fifth dimension in Sec.\re{probdens}, and to the extension to a massive particle 
propagating in the bulk
in Sec.\re{nonzero}.

For a few sets of boundary conditions (see Sec.\re{illustration}),
illustrative numerical evaluations are presented
to visualize the structure of some of the Kaluza-Klein mass eigenvalue towers.

To summarize, our model predicts the existence of mass state towers 
which could be observed 
at high energy colliders. 
The observation of a mass tower with 
its own specific carateristics
would validate the model.

%\newpage

\section{The five-dimensional metric. Allowed metric singularities \label{metricsec}}

We assume that
the warped five-dimensional space
with coordinates $x^A$ ($A=0,1,2,3,5$)
is composed of a flat
infinite four-dimensional subspace 
labeled by $x^{\mu}$ ($\mu=0,1,2,3$)
with signature ${\rm{diag}}(\eta_{\mu\nu})=(+1,-1,-1,-1)$ 
(underlying $SO(1,3)$ invariance) and a
spacelike fifth dimension with
coordinate $x^5\equiv s$ compactified
on the finite strip $0 \leq s\leq 2\pi R$. 

The most general non singular metric solution of Einstein's equations with
a stress-energy tensor identically zero and
a bulk negative cosmological constant $\Lambda$ is then locally, 
up to an overall metric rescaling,
\begin{equation}
dS^2=g_{AB}\,dx^A dx^B=H e^{-2{\tilde{\epsilon}} ks}\, \eta_{\mu\nu}\,dx^{\mu}dx^{\nu}-ds^2
\label{metric}
\end{equation}
\begin{itemize}

\item
the positive constant $k$ is related to
$\Lambda$ by
\begin{equation}
k=\sqrt{-\frac{\Lambda}{6}} \ > 0
\label{klambda}
\end{equation}

\item
${\tilde{\epsilon}}$ is an arbitrary sign

\item
$H$ is an arbitrary constant.

\end{itemize}

As stated in \cite{GN2}, it may be assumed 
that in some non necessarily connected region $S_{+}$ of $s$, 
${\tilde{\epsilon}}$ is ${+}1$ while in the 
complementary region $S_{-}$ ($S_{+}\cup S_{-}=[0,2\pi R]$)
it is ${-}1$. 
For physical reasons,
the metric must obviously be continuous. 
Hence, in a connected region with a given ${\tilde{\epsilon}}$ sign, the related
$H$ has to be constant throughout that region.  
For two regions with opposite signs of ${\tilde{\epsilon}}$,
joining at  
what we call a singularity point $s_s$, 
the continuity condition implies that the metric takes the following form in the
vicinity of that point
\bea
{\rm{for\ }} s<s_s&\quad& dS^2=\ Ce^{-2\epsilon k(s-s_s)}\, \eta_{\mu\nu}\, dx^{\mu}dx^{\nu}-ds^2
     \nonumber\\
{\rm{for\ }} s=s_s&\quad& dS^2=\ \ \quad C\ \, \eta_{\mu\nu}\,dx^{\mu}dx^{\nu}\quad\quad\ - ds^2
     \nonumber\\
{\rm{for\ }} s>s_s&\quad& dS^2=\ Ce^{2\epsilon k(s-s_s)} \, \eta_{\mu\nu}\,dx^{\mu}dx^{\nu}\ -ds^2
\label{singular}
\eea
with constant $C$ and a sign $\epsilon$.
The metric components $g_{\mu\mu}$ are continuous at $s=s_s$ as they should 
be, but
their first derivatives have a discontinuity $4 \epsilon  kC$ and
their second derivatives a $\delta$-function behavior.
In principle, there could be any finite number $N$ of such singularities.
As will be shown in  Sec.\re{circle},
if the number of singularities is even, the strip 
can in certain cases be closed into a circle. 

%\newpage

\section{A single metric singularity. Riemann equation. Hermiticity. 
Kaluza-Klein reduction. 
Boundary conditions. Solutions
\label{General}}

In this section, we restrict ourselves to a general discussion 
when there is a single metric singularity situated at $s=s_1$ on the finite
strip $0<s_1< 2\pi R$.
Then according to \re{singular} the metric \re{metric} is
\bea
&{\rm{for\ }} 0\leq s \leq s_1\quad& dS^2=e^{-2\epsilon k s}\, \eta_{\mu\nu}\,  dx^{\mu}dx^{\nu}-ds^2
     \nonumber\\
&{\rm{for\ }} s_1\leq s\leq 2\pi R\quad& dS^2=e^{2\epsilon k(s-2s_1)} \, \eta_{\mu\nu}\, dx^{\mu}dx^{\nu}-ds^2 
\label{onesingular}
\eea
where, without loss of generality, $C$ has been taken equal to $e^{-2\epsilon k s_1}$.

\subsection{Single metric singularity. Riemann equation \label{eqmotion}}

For complex scalar fields $\Phi(x,t)$ in a five-dimensional 
Riemann space with a compactified fifth dimension, 
the invariant scalar product is given by
\begin{equation}
\bigl(\Psi,\Phi\bigr)
=\int_{-\infty}^{+\infty} d^4x \int_0^{2\pi R}
ds\ \ \sqrt{g} \   \Psi^{*}(x,s)\,\Phi(x,s)\ .
    \label{scalprodgen}
\end{equation}

As discussed in App.\re{least}, the invariant equation of motion 
resulting from a least action principle applied to the action
\begin{equation}
{\cal{A}}
=\int_{-\infty}^{+\infty} d^4x\, \int_0^{2\pi R}ds
\ \,(\partial_A\Phi^*)\,\sqrt{g}\,g^{AB}\,(\partial_B \Phi)
\label{actiona}
\end{equation}
with vanishing field variations at the boundaries is
\begin{equation}
\square_{\rm{Riemann}}\Phi\equiv
\frac{1}{\sqrt{g}}\partial_A \sqrt{g}g^{AB}\partial_B\Phi =0\ .
\label{dalemb5}
\end{equation}
Away from $s=s_1$, the equation has no singularity.
For a massive scalar field in the bulk, see Sec.\re{nonzero}.

From the metric \re{onesingular}, $\sqrt{g}$ depends on $s$
\bea
&{\rm{for\ }} 0\leq s\leq s_1\quad& \sqrt{g}= e^{-4\epsilon k s}
     \nonumber\\
&{\rm{for\ }} s_1\leq s\leq 2\pi R\quad&  \sqrt{g}=e^{4 \epsilon  k(s-2 s_1)} 
\label{determ}
\eea
and is continuous as it should. 
The Riemann equation \re{dalemb5} then becomes
\bea
&&{\rm{for\ }} 0\leq s<s_1\hspace{0.5 cm}
     \nonumber\\
&& \quad\quad (e^{2\epsilon k s}\square_4
                                 -e^{4\epsilon k s}\partial_s e^{-4 \epsilon k s}\partial_s)
                                 \Phi(x^{\mu},s)=0  
     \label{basiceq1}\\
     \nonumber\\
&&{\rm{for\ }} s_1<s\leq 2\pi R
     \nonumber\\
&&\quad\quad (e^{-2\epsilon k(s-2s_1)}\square_4
                     -e^{-4\epsilon k(s-2s_1)}\partial_s e^{4 \epsilon k(s-2s_1)}\partial_s)
                                 \Phi(x^{\mu},s)=0
\label{basiceq2}
\eea
where $\square_4=\, \eta_{\mu\nu}\, \partial^{\mu}\partial^{\nu}$
is the usual four-dimensional d'Alembertian operator.

%\newpage

\subsection{Single metric singularity. Generalized hermiticity conditions {\label{herm}}} 

Following closely the discussion of our previous article \cite{GN2} dealing 
with Kaluza-Klein towers in warped spaces without metric singularities, 
we summarize and collect here the results 
which are valid for this extended case. 

%\vskip 0.2 cm
% \noindent ${\bullet}$ 
Remember that an operator $A$ is symmetric for a scalar product if
\begin{equation}
\bigl(\Psi,A\Phi\bigr)
=\bigl(A\Psi,\Phi\bigr)
\label{symmetric}
\end{equation}
for all the vectors $\Psi\in D(A)$ and $\Phi\in D(A)$,
i.e. if the adjoint operator $A^{\dagger}$
of the operator $A$ is an extension of $A$:
$A^{\dagger}\Phi=A\Phi$ for all $\Phi\in D(A)$
and $D(A^{\dagger})\supset D(A)$.
It is self-adjoint if $A^{\dagger}\Phi=A\Phi$ for all $\Phi\in D(A)$ and
moreover $D(A^{\dagger})=D(A)$,
i.e. if the operator is symmetric and if the equation \re{symmetric}
cannot be extended naturally to vectors $\Psi$ outside $D(A)$.

%\vskip 0.2 cm
%\noindent ${\bullet}$ 
One can easily check that
the operator $\square_{\rm{Riemann}}$ in \re{dalemb5} is formally symmetric,
by which we mean that it is symmetric up to boundary conditions.
Integrating twice by parts the symmetry equation 
\begin{equation}
\bigl(\Psi,\square_{\rm{Riemann}}\Phi\bigr)
=\bigl(\square_{\rm{Riemann}}\Psi,\Phi\bigr)
\label{symmetricRiem}
\end{equation}
for the scalar product \re{scalprodgen}, one finds
that the part of the operator in \re{basiceq1},\re{basiceq2} which is proportional to $\square_4$,
namely the operator defined by
\begin{equation}
A_1\equiv\left\{
\begin{array}{lcl}
{\rm{for\ }} 0\leq s\leq s_1
&:& e^{2\epsilon k s}\square_4
     \nonumber\\
{\rm{for\ }} s_1\leq s\leq 2\pi R
&:& e^{-2\epsilon k(s-2s_1)}\square_4
\end{array}
\right\}\ ,
\label{opA2}
\end{equation}
is fully symmetric. 
The second part in \re{basiceq1},\re{basiceq2} (involving derivatives with respect to $s$), namely
the operator $A_2$ defined by
\begin{equation}
A_2\equiv\left\{
\begin{array}{lcl}
{\rm{for\ }} 0\leq s<s_1
&:& e^{4\epsilon k s}\partial_s e^{-4 \epsilon k s}\partial_s
     \nonumber\\
{\rm{for\ }} s_1<s\leq 2\pi R
&:& e^{-4\epsilon k(s-2 s_1)}\partial_s e^{4 \epsilon k(s-2 s_1)}\partial_s
\end{array}
\right\}
\label{opA3}
\end{equation}
is formally symmetric. The condition of full symmetry of $A_2$ is expressed by
the boundary relation which is the $x^{\mu}$ integral of
\bea
&\lim_{\eta\rightarrow 0^+}\Biggl\{
     &\Bigl[e^{-4\epsilon k s} \Bigl(\Psi^*(\partial_s \Phi)
     -(\partial_s \Psi^*) \Phi\Bigl)\Bigr]
     \rule[- 4 mm]{0.1 mm}{10 mm}_{\ 0}^{\ s_1-\eta}\Biggr.
     \nonumber\\
&&\quad\quad\Biggl.+\Bigl[e^{4\epsilon k(s-2s_1)}\Bigl(\Psi^*(\partial_s \Phi)
     -(\partial_s \Psi^*) \Phi\Bigl)\Bigr]
     \rule[- 3 mm]{0.12 mm}{10 mm}_{\ s_1+\eta}^{\ 2\pi R}\Biggr\}=0\ .
\label{BCwarpsing}
\eea

%\vskip 0.2 cm
%\noindent ${\bullet}$ 
Unfortunately, the operators $A_1$ and $A_2$ do not commute and 
hence cannot be diagonalized together. 
Multiplying on the left the equation \re{basiceq1} 
by $e^{-2\epsilon k s}$ and \re{basiceq2} by
$e^{2\epsilon k(s-2s_1)}$, one obtains the following operators
\begin{equation}
 B_1\equiv\Biggl\{\quad{\rm{for\ }} 0\leq s \leq 2\pi R
\quad:\quad\square_4\Biggr\}
\label{opB1}
\end{equation}
and  
\begin{equation}
B_2\equiv\left\{
\begin{array}{lcl}
{\rm{for\ }} 0\leq s<s_1&
:& e^{2\epsilon k s}\partial_s e^{-4\epsilon  k s}\partial_s
     \nonumber\\
{\rm{for\ }} s_1<s\leq 2\pi R&
:& e^{-2\epsilon k(s-2s_1)}\partial_s e^{4 \epsilon k(s-2s_1)}\partial_s
\end{array}
\right\}\ .
\label{opB2}
\end{equation}
The operators $B_1$ and $B_2$ commute and can be diagonalized together allowing the
interpretation of the eigenvalues of $B_2$, if they are real, in terms of masses squared.

%\vskip 0.2 cm
%\noindent ${\bullet}$ 
However,
as discussed at length in \cite{GN2},
the operator $B_2$ is not even formally symmetric for the scalar product \re{scalprodgen}.
We showed that by a suitable non unitary change of basis
\bea
{\widetilde{B}}_2&=&V B_2 V^{-1}
        \nonumber\\
{\widetilde{\Phi}}&=&V\Phi
\label{warpchange}
\eea
defined here by the continuous function $V$
\begin{equation}
V\equiv\left\{
\begin{array}{c}
{\rm{for\ }} 0\leq s\leq s_1
\quad\quad\quad e^{\epsilon k s}
     \nonumber\\
{\rm{for\ }} s_1\leq s\leq 2\pi R
\quad\quad e^{-\epsilon k(s-2s_1)}
\end{array}
\right.
\label{Vform}
\end{equation}
the operator ${\widetilde{B}}_2$ happens to be formally symmetric for the 
scalar product deduced from \re{scalprodgen} and \re{warpchange}, namely
\bea
\bigl({\widetilde{\Psi}},{\widetilde{\Phi}}\bigr)
&=&\int_{-\infty}^{+\infty} d^4x \int_0^{s_1}
ds\ e^{-6 \epsilon  k s}\   {\widetilde{\Psi}}^{*}\,{\widetilde{\Phi}} 
    \nonumber\\
& & +\int_{-\infty}^{+\infty} d^4x \int_{s_1}^{2\pi R}
ds\  e^{6 \epsilon  k(s-2s_1)}\   {\widetilde{\Psi}}^{*}\,{\widetilde{\Phi}} \ .
\label{scalprodtilde}
\eea
We found that
the ${\widetilde{\Psi}}$ boundary relation
arising from the requirement that $B_2$ 
be symmetric for the scalar product \re{scalprodtilde}
turns out to be exactly equal to the boundary relation 
\re{BCwarpsing}
for the untransformed field $\Psi$ when requiring symmetry of $A_2$.

Thus, even though the operator $B_2$ is not even formally symmetric, 
it is equivalent by a non unitary transformation to a formally symmetric operator.  
Once the correct boundary conditions satisfying the boundary relation \re{BCwarpsing}
are imposed, thereby defining the Hilbert space of the field solutions, the operator 
$B_2$ becomes fully symmetric 
and its eigenvalues are real. 
This can be brought in parallel with the recently discovered examples of real eigenvalues for non 
hermitian operators \cite{Ben}, \cite{FN}.  

The boundary conditions resulting from the boundary relation \re{BCwarpsing} 
are analyzed in Sec.\re{HBC}.

%\newpage

\subsection{Single metric singularity. The Kaluza-Klein reduction equations 
and the mass eigenvalue equations {\label{KKreduction}}}

We adopt the usual Kaluza-Klein reduction \cite{KK}
with separation of the variables $x^{\mu}$ and $s$
for the real massless scalar field $\Phi(x,t)$ 
\be
\Phi(x^{\mu},s)=\sum_{n}\phi^{[x]}_n(x^{\mu})\,\phi^{[s]}_n(s)\ .
\label{KKreduc}
\end{equation}
The field $\Phi(x^{\mu},s)$ is a solution of the Riemann equations
\re{basiceq1},\re{basiceq2} written in terms of
the operators $B1$ \re{opB1} and $B2$ \re{opB2} 
\begin{equation}
(B_1-B_2)\Phi(x^{\mu},s)=0
\label{masszeroeq}
\end{equation} 
if
\begin{equation}
B_1\ \phi^{[x]}_n(x^{\mu})=-m_n^2\ \phi^{[x]}_n(x^{\mu})
      \label{B1eq}
\end{equation}
and if
\begin{equation}      
B_2\ \phi^{[s]}_n(s)=-m_n^2\phi_n^{[s]}(s)\ .
       \label{B2eq}
\end{equation}
In any four-dimensional brane, if these $m_n^2$ eigenvalues are real, 
positive $m_n^2$ will correspond to scalar
particles, negative  $m_n^2$ to scalar tachyons and 
$m_n^2=0$ to zero mass scalars. 
We proved in the Sec.\re{herm} that 
by imposing the boundary relation \re{BCwarpsing}
the eigenvalues of $B_2$ are effectively real.
As will be discussed later (Sec.\re{physicmass}), 
the observable physical masses 
derive from the eigenvalue masses in a way 
depending on the position of the brane  
on the $s$ strip.

%\newpage

\subsection{Single metric singularity. General formulation of the boundary conditions
\label{HBC}}

In this Section, we derive from the boundary relation \re{BCwarpsing} 
the most general boundary conditions to be imposed on
the Kaluza-Klein reduced fields 
$\phi^{[s]}_n(s)$ from Eq.\re{KKreduc}, in the case of
a single metric singularity at $s = s_1$.
Using the following notations
\bea
\phantom{\Biggl\{\Biggr\}}
      \phi_0&=&\phi^{[s]}_n(0)
      \nonumber\\
\phantom{\Biggl\{\Biggr\}}
      \partial\phi_0&=&(\partial_s\phi^{[s]}_n)(0)
      \nonumber\\      
\phantom{\Biggl\{\Biggr\}}
      \phi_l&=&e^{-2\epsilon k s_1}\lim_{\eta\rightarrow 0^+}\phi^{[s]}_n(s_1-\eta)
      \nonumber\\
\phantom{\Biggl\{\Biggr\}}
      \partial\phi_l&=&e^{-2\epsilon k s_1}\lim_{\eta\rightarrow 0^+}(\partial_s\phi^{[s]}_n)(s_1-\eta)
      \nonumber\\      
\phantom{\Biggl\{\Biggr\}}
      \phi_r&=&e^{-2\epsilon k s_1}\lim_{\eta\rightarrow 0^+}\phi^{[s]}_n(s_1+\eta)
      \nonumber\\
\phantom{\Biggl\{\Biggr\}}
      \partial\phi_r&=&e^{-2\epsilon k s_1}\lim_{\eta\rightarrow 0^+}(\partial_s\phi^{[s]}_n)(s_1+\eta)
      \nonumber\\      
\phantom{\Biggl\{\Biggr\}}
      \phi_R&=&e^{4\epsilon (\pi R -s_1)}\phi^{[s]}_n(R)
      \nonumber\\
\phantom{\Biggl\{\Biggr\}}
      \partial\phi_R&=&e^{4\epsilon (\pi R -s_1)}(\partial_s\phi^{[s]}_n)(0)
\label{appnotation}     
\eea      
and similarly for $\psi$ (related to $\psi_p^{[s]}(s)$), the basic boundary relation \re{BCwarpsing} 
becomes after the Kaluza-Klein reduction \re{KKreduc}
\bea
(\psi_R^*\,\partial\phi_R-\partial\psi_R^*\,\phi_R)
-(\psi_r^*\,\partial\phi_r-\partial\psi_r^*\,\phi_r)
 \quad      &&
         \nonumber\\
+(\psi_l^*\,\partial\phi_l-\partial\psi_l^*\,\phi_l)
-(\psi_0^*\,\partial\phi_0-\partial\psi_0^*\,\phi_0)&=&0\ .
\label{BCwarpsing2}
\eea
This boundary relation implies that there must be exactly four boundary
conditions, expressed by four independent linear relations between the eight
components of the vector
\begin{equation}
\Phi=\left(
     \matrix{
           \phi_0\cr
           \partial\phi_0\cr
            \phi_l\cr
           \partial\phi_l\cr
           \phi_r\cr
           \partial\phi_r\cr
           \phi_R\cr
           \partial\phi_R\cr
           }
\right)\ .
\label{vectorphi}
\end{equation}
The same boundary conditions must hold true for the corresponding vector
$\Psi$. In terms of $\Phi$ and $\Psi$, the boundary relation \re{BCwarpsing2}
is written in matrix form
\begin{equation}
\Psi^+ S^{[8]}\Phi =0
\label{boundrel}
\end{equation}
with the $8 \times 8$ antisymmetric matrix $S^{[8]}$
\begin{equation}
S^{[8]}=\left(
     \matrix{
           S^{[4]} & 0^{[4]}\cr
           0^{[4]} & S^{[4]}
           }
\right)=1^{[4]}\otimes S^{[4]}
\label{matrixS8}
\end{equation}
and
\begin{equation}
S^{[4]}=\left(
     \matrix{
           i\sigma_2 & 0^{[2]}\cr
           0^{[2]} & -i\sigma_2 
           }
\right)=\sigma_3\otimes (i\sigma_2)\ .
\label{matrixS4}
\end{equation}

The four boundary conditions are expressible in terms of a $4 \times 8$ matrix
$M$ of rank 4 as
\begin{equation}
M\Phi=0\ .
\label{BCN}
\end{equation}
For any $M$, a permutation $P$ can be chosen such that these four boundary
conditions are equivalent to
\begin{equation}
P\Phi=V_P^{[8]}P\Phi
\label{BCV}
\end{equation}
with the $8\times 8$ matrix $V^{[8]}_P$, written in terms of a $4\times 4$ matrix $V^{[4]}_P$
(depending on $P$) and the unit matrix $1^{[4]}$,
\begin{equation}
V^{[8]}_P=\left(
     \matrix{
           1^{[4]}   & 0^{[4]}\cr
           V^{[4]}_P & 0^{[4]} 
           }
\right)\ .
\label{matrixV8}
\end{equation}
Writing $\Phi_P \equiv P\Phi$ in terms of its four upper elements $\Phi_P^u$
and its four down elements $\Phi_P^d$
\begin{equation}
\Phi_P=\left(
     \matrix{
           \Phi_P^u\cr
          \Phi_P^d 
           }
\right)
\label{vectorphiP}
\end{equation}
one finds that the four first equations in \re{BCV} are trivial while the four last
equations express the boundary conditions equivalent to \re{BCN}
\begin{equation}
\Phi_P^d=V_P^{[4]}\Phi_P^u\ .
\label{BCcaseP}
\end{equation}
This is in agreement with the observation that, from \re{BCN}, 
there exists always a permutation $P$ of the components of $\Phi$
such that four components ($\Phi_P^d$) are linear functions 
of the four other independent components ($\Phi_P^u$).

Writing $S^{[8]}_P$ the transformed of $S^{[8]}$ under the permutation $P$
\begin{equation}
S_P^{[8]}=P S^{[8]} P^{-1}
\label{SP1}
\end{equation}
the matrix $V^{[8]}_P$ expressing the allowed boundary conditions \re{BCV}
must satisfy the matrix equation
\begin{equation}
V_P^{[8]+}\,S_P^{[8]}\,V^{[8]}_P=0\ .
\label{BCP}
\end{equation}
This follows from the fact that the boundary relation \re{boundrel} 
then depends on $\Phi^u$ and 
$\Psi^{u+}$ only, which are arbitrary.

With the four $4\times 4$ matrices $S_{Pj},j = 1,\dots,4$ defined from $S_P^{[8]}$ as
\begin{equation}
S_P^{[8]}=\left(
     \matrix{
           S_{P1}^{[4]} & S_{P2}^{[4]} \cr
          S_{P3}^{[4]} & S_{P4}^{[4]} 
           }
\right)\ ,
\label{SP2}
\end{equation}
the boundary relation \re{BCP} leads explicitly to an equation for $V^{[4]}_P$
\begin{equation}
S_{P1}^{[4]} + V_P^{[4]+}\,S_{P3}^{[4]} + S_{P2}^{[4]}\,V^{[4]}_P + V_P^{[4]+}\,S_{P4}^{[4]}\,V^{[4]}_P=0\ .
\label{SPBC}
\end{equation}

It should be stressed that different choices of $P$ may lead to equivalent
boundary conditions, in particular, by multiplying a given $P$ by further permutations
within the four elements of $\Phi^u_P$ or within the four elements of $\Phi^d_P$.

A few examples of sets of boundary conditions are given in App.\re{appex}.

%\newpage 

\subsection{Single metric singularity. Solutions for the fields and
mass eigenvalues {\label{solutionform}}}

For positive $m_n^2$, the solutions of \re{B2eq},\re{opB2} 
are linear superpositions of the Bessel functions 
$J_2$ and $Y_2$ on the left side $[L]$ as well as 
on the right side $[R]$ of the singular point
\bea
&{\rm{for\ }}0\leq s<s_1\hspace{1.5 cm}&
     \label{solwarpKK1}
     \\
&\phi^{[s]}_n(s)=e^{2\epsilon k s}
               \Biggl(\sigma_{n}^{[L]} 
                             J_2\left(\frac{m_n e^{\epsilon k s)}}{k}\right)                              
                            +\tau_{n}^{[L]} 
                             Y _2\left(\frac{m_n e^{\epsilon k s}}{k}\right)\Biggr)&
         \nonumber\\
&{\rm{for\ }}s_1<s\leq 2\pi R \hspace{1 cm}&
     \label{solwarpKK2}
     \\
&\phi^{[s]}_n(s)=e^{-2\epsilon k(s-2s_1)}
                  \Biggl(\sigma_{n}^{[R]} 
                             J_2\left(\frac{m_n e^{-\epsilon k(s-2 s_1)}}{k}\right)\Biggr.             
                           +\tau_{n}^{[R]} 
                             Y _2\left(\frac{m_n e^{-\epsilon k(s-2 s_1)}}{k}\right)\Biggr)&
\nonumber
\eea
where $\sigma_{n}^{[L]},\tau_{n}^{[L]},\sigma_{n}^{[R]},\tau_{n}^{[R]}$ 
are four arbitrary integration constants. 
In general, the boundary conditions 
\re{BCN} or equivalently \re{BCcaseP} provide
four linear homogeneous relations among the four integration constants.
In order to have a non trivial solution for the arbitrary constants, 
the related $4\times 4$ matrix 
must be of rank three and hence
must have a zero determinant. This leads to an equation for 
$m_n$ which determines the mass eigenvalues 
building up the Kaluza-Klein tower.

In some cases there exists a scalar zero mass state in the tower. 
The solution takes then the special form
\bea
{\rm{for\ }}0\leq s<s_1&&\phi^{[s]}_0(s)
    =\sigma_{0}^{[L]}e^{4\epsilon k s}
    +\tau_{0}^{[L]} \ .
         \nonumber\\
{\rm{for\ }}s_1<s\leq 2\pi R&&\phi^{[s]}_0(s)
    =\sigma_{0}^{[R]}e^{-4\epsilon k(s-2 s_1)}
     +\tau_{0}^{[R]}  \ .
     \label{solwarpKK0}
\eea
These zero mass states occur only for specific boundary condition
parameters. They are worth the dedicated 
Sec.\re{zeromassBC}.

In some cases there exists a scalar tachyon in the tower
corresponding to a negative $m_n^2=-h^2<0$ eigenvalue of \re{B2eq},\re{opB2}. 
The solution is then a superposition 
of the modified Bessel functions $I_2$ and $K_2$ 
\bea
&{\rm{for\ }}0\leq s<s_1&
         \label{solwarpKK3}\\
&\phi^{[s]}_t(s)=e^{2\epsilon k s}
\Biggl(\sigma_{t}^{[L]} 
                        I_2\left(\frac{h\, e^{\epsilon k s}}{k}\right)
                              +\tau_{t}^{[L]} 
                        K _2\left(\frac{h\, e^{\epsilon k s}}{k}\right)\Biggr)&
\nonumber\\
&{\rm{for\ }}s_1<s\leq 2\pi R\hspace{1 cm}&
         \label{solwarpKK4}\\
&\phi^{[s]}_t(s)=e^{-2\epsilon k(s-2s_1)}
\Biggl(\sigma_{t}^{[R]} 
                        I_2\left(\frac{h\, e^{-\epsilon k(s-2s_1)}}{k}\right)
                              +\tau_{t}^{[R]} 
                        K _2\left(\frac{h\, e^{-\epsilon k(s-2s_1)}}{k}\right)\Biggr)\ .&
   \nonumber
\eea
The boundary conditions \re{BCN},\re{BCcaseP}
imply that the four integration constants 
$\sigma_{t}^{[L]}$, $\tau_{t}^{[L]}$, $\sigma_{t}^{[R]}$, $\tau_{t}^{[R]}$
satisfy four linear homogeous relations.
The mass eigenvalues corresponding to the tachyon states 
are obtained by imposing again that 
the related determinant is zero. 
Solutions for these usually lonely states occur 
only in certain ranges of the boundary conditions parameters.

%\newpage

\subsection{Single metric singularity. Specific zero mass conditions {\label{zeromassBC}}}

If there is zero mass state in the tower, the boundary conditions 
lead as before to four linear homogeneous relations 
among the four integration constants 
$\sigma_{0}^{[L]}$, $\tau_{0}^{[L]}$, $\sigma_{0}^{[R]}$, $\tau_{0}^{[R]}$ of Eq.\re{solwarpKK0}.
The condition that the related determinant is zero implies, for a zero mass state to exist, 
a  constraint
between the boundary condition parameters 
and the parameters $k,R,s_1$.

In certain  
cases, the above matrix can also be of rank two (or lower) rather than three
if additional relations involving the parameters of the boundary conditions
and the parameters $k,R,s_1$ are satisfied
In this situation there exist two (or more) linearly independent solutions 
and hence a doubly (or higher) degenerated zero mass. 

In the case of a single zero mass state,
the parameter constraint equation defines a surface 
in the parameter space. In general, if one follows a path in the parameter
space which crosses the constraint surface, there is
tower for each set of parameters. On
one side of the surface, the tower has a lowest mass 
eigenvalue which goes smoothly toward zero, takes the value zero as the path goes through 
the surface and emerges as a tachyon state with low $h^2=-m^2$ on the other side
(see for example Table\re{tableKK195}).

%\newpage

\section{$N$ metric singularities. Riemann equation. Hermiticity.
Kaluza -Klein reduction. 
Closure into a circle \label{anysing}}

The extension of the preceding 
to a warped space 
with an arbitrary number $N$ of metric singularities situated at the points
$0<s_1<s_2,\dots,s_N<2\pi R$ on the strip is straightforward. 
There are $N+1$ intervals $I_i,\ i=0,\dots,N$ 
\begin{equation} 
I_0=[0,s_1],\ I_1=[s_1,s_2],\ \dots\ ,\ I_{N-1}
=[s_{N{-}1},s_N],\ I_N=[s_N,2\pi R] 
\label{intervals}
\end{equation}
of respective length
\begin{equation}
l_0=s_1,\,l_1=s_2{-}s_1,\,l_2=s_3{-}s_2,\,\dots,\,l_N=2\pi R{-}s_N \ .
\label{lengths}
\end{equation}
Defining
\begin{equation} 
r_i=-2(-1)^{i{+}1}\left(\sum_{j=0}^{i-1}(-1)^j s_{i-j}\right)
    \label{theri1}
\end{equation}
(note $r_0 {=} 0$) equivalent to    
\bea
r_{2i}&=&2\sum_{j=1}^{i} l_{2j{-}1}
    \nonumber\\
r_{2i+1}&=&-2\sum_{j=0}^{i} l_{2j}  \ ,
\label{theri2}
\eea
the metric takes the form  
\begin{equation}
{\rm{for\ }}s\in I_i \quad :\quad dS^2
=e^{-2k\epsilon \left((-1)^{i}s-r_i\right)} dx_{\mu}dx^{\mu}-ds^2\quad (i=0,\dots,N) \ .
\label{manysing}     
\end{equation}
Without loss of generality, since $r_0=0$, 
the coefficient $H$ (\re{metric}) 
has been adjusted to one in the first interval $I_0$. 
The sign of the coefficient of $s$ in the exponent alternates 
between $\epsilon$ and $-\epsilon$ for the intervals $I_i$ 
with even and odd $i$. 
The end points of each interval 
are thus singular points, except $s=0$ and $s=2\pi R$ 
(see however the special case in Sec.\re{circle}).

%\newpage

\subsection{$N$ metric singularities. Hermiticity and boundary conditions {\label{hermN}}}

If there are $N>1$ singularities,
the generalization of the boundary relation \re{BCwarpsing2}
and of the allowed boundary conditions as introduced in Sec.\re{HBC}
is straightforward. There are $2N+2$ boundary edges: 
the $N$ left edges and the $N$ right edges of the intervals
\re{intervals} together with the edges 
$0$ and $2\pi R$ of the $s$-domain . The vector $\Phi$ generalizing \re{vectorphi}
has $4N+4$ components and the matrix $M$ \re{BCN} expressing the boundary conditions 
is a $(2N{+}2)\times (4N{+}4)$ matrix of rank $2N{+}2$. The matrix $S^{[4N+4]}$ which expresses
the boundary relation generalizing \re{boundrel}
is a block diagonal antisymmetric
matrix made of $N+1$ matrices $S^{[4]}$ \re{matrixS4}. A permutation exists such that the formulae 
\re{BCV}, \re{SP1} and \re{BCP} hold true with the index $[8]$ replaced by $[4N{+}4]$,  
in particular
\begin{equation}
V^{[4N{+}4]}_P=\left(
     \matrix{
           1^{[2N{+}2]}   & 0^{[2N{+}2]}\cr
           V^{[2N{+}2]}_P & 0^{[2N{+}2]} 
           }
\right)\ .
\label{matrixVN}
\end{equation}
In \re{vectorphiP}, $\Phi_P^u$ is composed of the $2N{+}2$ up elements of $\Phi_P$
while $\Phi_P^d$ is composed of the $2N{+}2$ down elements. 
The generalisation of \re{BCcaseP}, of  \re{SP2} and of \re{SPBC} is then straightforward.
One has 
\begin{equation}
\Phi_P^d=V_P^{[2N{+}2]}\Phi_P^u
\label{BCcasePN}
\end{equation}
as well as
\begin{equation}
S_P^{[2N{+}2]}=\left(
     \matrix{
          S_{P1}^{[2N{+}2]} & S_{P2}^{[2N{+}2]} \cr
          S_{P3}^{[2N{+}2]} & S_{P4}^{[2N{+}2]} 
           }
\right)
\label{SP2N}
\end{equation}
and
\begin{equation}
S_{P1}^{[2N{+}2]} + V_P^{[2N{+}2]+}\,S_{P3}^{[2N{+}2]} 
+ S_{P2}^{[2N{+}2]}\,V^{[2N{+}2]}_P + V_P^{[2N{+}2]+}\,S_{P4}^{[2N{+}2]}\,V^{[2N{+}2]}_P=0\ .
\label{SPBCN}
\end{equation}
With the restrictions on the boundary parameters in $V^{[2N{+}2]}_P$ arising from \re{SPBCN}, 
the equations \re{BCcasePN} express the allowed $2N{+}2$ boundary conditions
as the generalisation of the equations \re{BCcaseP}, \re{SP2}, \re{SPBC}.

%\newpage

\subsection{$N$ metric singularities. Riemann equation. Solutions. Mass eigenvalues \label{anysingeigen}}

Following closely the discussion of the case 
with a single singularity (Sec.\re{illustration}), 
the Kaluza-Klein reduction equations \re{B1eq},\re{B2eq} 
for a real massless scalar field lead to the following equations: 

$\bullet${\hspace{0.2 cm}} for $\phi^{[x]}_n(x^{\mu})$, one has
\begin{equation}
       \square_4\  \phi^{[x]}_n(x^{\mu})
                  =-m_n^2\ \phi^{[x]}_n(x^{\mu}) 
\label{B1eqsing}
\end{equation}

$\bullet$ {\hspace{0.2 cm}}for $\phi^{[s]}_n(s)$, the equation depends on the interval $I_i$ \re{intervals}, \re{theri1}
\begin{equation}
e^{2\epsilon k\left(\left({-}1\right)^i s-r_i\right)}\partial_s 
e^{-4\epsilon  k\left(\left({-}1\right)^i s-r_i\right)}\partial_s\phi_n^{[s]}(s)
=-m_n^2\phi_n^{[s]}(s)\ .
\label{B1eqsing2}
\end{equation}
 
The form of the solution for $\phi^{[s]}_n(s)$ depends both on the intervals $I_i$ 
and on the sign of the eigenvalue $m_n^2$: 
\bea
&{\hspace{-8 cm}}\bullet{\hspace{0.2 cm}}{\rm{for\ }}s\in I_i{\rm{\ and\ }}m_n^2 >0&
         \label{manysingsolu1}\\
  &\phi^{[s]}_n(s)=e^{2\epsilon k\left(\left({-}1\right)^i s-r_i\right)}
  \Biggl(\sigma_{n}^{[i]} 
           J_2\left(\frac{m_n e^{\epsilon k
           \left(\left({-}1\right)^i s-r_i\right)}}{k}\right)           
                                +\tau_{n}^{[i]} 
           Y_2\left(\frac{m_n e^{\epsilon k
           \left(\left({-}1\right)^i s-r_i\right)}}{k}\right)
                                      \Biggr)&
    \nonumber\\
&{\hspace{-8 cm}}\bullet{\hspace{0.2 cm}}{\rm{for\ }}s\in I_i{\rm{\ and\ }}m_n^2 =0&
         \label{manysingsolu2}\\
  &{\hspace{-7.5 cm}}\Biggl.
  \phi^{[s]}_0(s)=\sigma_{0}^{[i]}e^{4\epsilon k
  \left(\left({-}1\right)^i s-r_i\right)}
                              +\tau_{0}^{[i]}\Biggr.&
         \nonumber\\
&{\hspace{-6.5 cm}}\bullet{\hspace{0.2 cm}}{\rm{for\ }}s\in I_i{\ \rm{\ and\ }}m_n^2=-h^2 <0&         
         \label{manysingsolu3}\\
  &\phi^{[s]}_h(s)=e^{2\epsilon k
  \left(\left({-}1\right)^i s-r_i\right)}
  \Biggl(\sigma_{t}^{[i]} 
            I_2\left(\frac{h\, e^{\epsilon k
            \left(\left({-}1\right)^i s-r_i\right)}}{k}\right)
                         +\tau_{t}^{[i]} 
            K_2\left(\frac{h\, e^{\epsilon k
            \left(\left({-}1\right)^i s-r_i\right)}}{k}\right)\Biggr)\,.&
         \nonumber
\eea         

There are altogether $(2N{+}2)$ integration constants $\sigma^{[i]},\tau^{[i]}$ 
which must satisfy 
$(2N{+}2)$ linear homogeneous relations resulting from the $2N{+}2$ boundary conditions  
\re{BCcasePN}. In order to obtain a non trivial solution for the integration constants
the related $(2N{+}2)\times (2N{+}2)$ 
determinant must vanish.
As in the case with one singularity ($N=1$), the 
condition that the determinant is zero provides either
the mass eigenvalue equation, or the zero mass constraint 
on the parameters
or the tachyon eigenvalue equation. 

%\newpage

\subsection{$N$ metric singularities. Closure into a circle \label{circle}}

Finally, the strip could be closed into a circle by identifying 
the points $s=0$ and $s=2\pi R$, 
with $R$ interpreted as the radius of the circle. 
For this to be the case, the following requirements must hold.

There must be at least one singularity. 
Indeed if there are none, the metric is given by \re{metric} 
throughout the strip and cannot be made 
identical for $s=0$ and $s=2\pi R$, 
in disagreement with the continuity requirement. 

By rotation around the circle, the first singularity 
can always be placed at the closing point. 
Hence, if $g_{\mu\mu}$ is decreasing at the right of $s=0$ ($\epsilon=1$), 
it must be increasing at the left of $s=2\pi R$ (inversely if $\epsilon=-1$). 
Since the sign of $s$ in the exponential \re{manysing} 
changes every time one crosses a singularity, there must be altogether 
an odd number $2p-1$ of singularities distinct 
from the one at the closure point. 
The total number of singularities must hence be even $2p>0$ 
and situated at the points $s_0=0,s_1,s_2,\dots,s_{2p-1}$.

For the metric to be continuous at the closure point, 
the total range where the sign of $s$ in the exponential 
is positive must be equal to the total range where it is negative 
and hence equal to one half of the total range $2\pi R$. 
Thus, with the lengths $l_i$ defined in \re{lengths},
we have
\begin{equation}
\sum_{j=0}^{j=p-1}l_{2j}=\sum_{j=0}^{j=p-1}l_{2j+1}=\pi R\ .
\label{ranges}
\end{equation}

%\newpage

\section{Physical considerations \label{physdis}}

\subsection{Physical discussion of the boundary conditions \label{physdis1}}

The most general sets
of allowed boundary conditions are given in
Sec.\re{hermN}. 
The physical meaning of these conditions is worth some discussion.

Indeed, they impose relations on the $2N{+}2$ values 
of the fields  and $2N{+}2$ values of their derivatives 
at the left and right sides of the singular points and at the edge points of the $s$-domain. 
This, at first sight, seems to mean that the field must explore 
at once its full domain. In other words, locality seems to be broken or an action at a distance 
appears to take place. Quantum mechanics is customary of this type of behavior. 
The most famous example is the Einstein-Podolski-Rosen paradox \cite{EPR}, the correlation between 
the spin orientations of  a pair of particles originating from the decay of a scalar particle.  
In our mind, this is a convincing argument for considering that 
our new boundary conditions are of physical relevance.

Nevertheless, in the numerical applications, we choose, 
rather arbitrarily, to limit ourselves to
more conventional and naive boundary conditions.
We select the subsets of boundary conditions 
such that the values of the fields and of their derivatives
at the two sides of any internal singularity are 
directly connected 
to each other, but neither to the values at the other
singularities nor to the values at the edges of the $s$-domain. In some sense, these
subsets satisfy the 
locality criterion, however not fully as the field has to test its values across the 
singularity. 

On the other hand, we maintain some
non-locality in admitting that the values of the fields (and of their derivatives) are 
possibly related 
from one edge ($s=0$) to the other edge ($s=2\pi R$) of the domain.

Generalizing equation \re{transition} of case $B$ of the App.\re{appex}, we take every
one (remark that this is an arbitrary choice) of 
the $N$ singularities to be either periodic ($\delta_i=1$) or antiperiodic ($\delta_i=-1$), so we relate
the values of the fields (and derivatives) on the left and on the right of any $i$-singularity by
\bea
\phi_r^i
           &=&\delta_i\, \phi_l^i
      \nonumber\\
\partial \phi_r^i
          &=&\delta_i\, \partial \phi_l^i\ .
\label{transitioni}
\eea  
For the conditions at the edges, we essentially take either \re{boundB2}, which for real fields is written
\begin{equation}
\left(
     \matrix{
           \phi_R\cr
           \partial \phi_R
           } \right)
         =
     \left(
     \matrix{
           \alpha'&\beta'\cr
           \gamma'&\delta'
           } \right)
           \left(
     \matrix{
           \phi_0\cr
           \partial \phi_0
           } \right)      
     \quad,\quad \alpha'\delta'-\beta'\gamma' =1\ ,
\label{boundB2N}
\end{equation}
corresponding to the lines $A1$ and $A2$ of Table\re{tableBCdeux},
or a case analogous to the diagonal subcase of case $C$ in App.\re{appex} (which are 
of Sturm Liouville types) 
\bea
\kappa_0\,\partial\phi_0&=&\rho_0\,\phi_0
     \nonumber\\
\kappa_R\,\partial\phi_R&=&\rho_R\,\phi_R
\label{boundCspe}     
\eea
corresponding to the lines $A3$, $A4$ and $A5$ of Table\re{tableBCdeux}.
Is should be noted that in this last case, the boundary conditions are fully local at the edges.  

Summarizing in the case of a single periodic or antiperiodic singularity, 
our choice of boundary conditions, compatible with the boundary relation \re{BCwarpsing2},
leads (including the trivial set $A6$) to the six
independent sets of Table\re{tableBCdeux}.

When the strip in closed onto itself by identifying the points $s=0$ and $s=2\pi R$, 
with a periodic or antiperiodic singularity located in the middle at $s_1=\pi R$, 
we will also consider that the closure point, which becomes a metric singularity, 
is periodic or antiperiodic  
\bea
\left\{
\begin{array}{rcl}
\phi_R&=&\delta_0\, \phi_0
    \\
\partial\phi_R
      &=&\delta_0\,\partial\phi_0   
    \end{array}
    \right.\ .
   \label{circleBC}  
\eea

%\newpage

\subsection{The high mass scale. The Planck scale \label{Plscale}}

Our basic assumption is that
there is only one high mass scale in the theory
that we will 
naturally assume to be the Planck mass
\begin{equation}
M_{{\rm{Pl}}}\approx 1.22\ 10^{16}\ {\rm{TeV}}\ ,
\label{Planck}
\end{equation}
although any other high mass scale would be adequate for our purpose.

Any dimensionfull parameter $p$ with energy dimension $d$ is of the order
\bea
p&=&{\overline{p}}\left(M_{{\rm{Pl}}}\right)^d
      \nonumber\\
{\overline{p}}&:&{\rm{a\ pure\ number\ of\ order\ one\,.}}      
\label{barpara}
\eea
In particular 
$k={\overline{k}}M_{{\rm{Pl}}}$ and $R={\overline{R}}\left(M_{{\rm{Pl}}}\right)^{-1}$.
The boundary condition parameters which have a energy dimension 
scale also with the Planck mass, as 
for example the parameters $\alpha_2$, $\alpha_3$, $\rho_1$ ... etc 
which appear in Table \re{tableBCdeux}. We call the assumption that ${\overline{p}}$ is neither
a large nor a small number the 
``one-mass-scale-only'' hypothesis. In particular, 
$kR\,=\,\overline{k}\overline{R}$ is one of the 
major parameters of the model which governs the reduction from the high mass scale to the 
TeV scale for the low lying masses in the towers.

%\newpage

\subsection{The Physical Masses \label{physicmass}}

For a four-dimensional observer supposed to be sitting at $s=s_{\rm{phys}}$
in a given $I_i$ interval \re{intervals},
the metric \re{manysing} 
\begin{equation}
dS^2=e^{-2\epsilon k((-1)^i s_{\rm{phys}}-r_i)}dx_{\mu} dx^{\mu}-ds^2
\label{canonicmetric}
\end{equation}
can be transformed 
in canonical form
\begin{equation}
dS^2=d{\widetilde{x}}_{\mu} d{\widetilde{x}}^{\mu}-ds^2
\label{canonicmetric2}
\end{equation}
by the following rescaling 
\begin{equation}
{\widetilde{x}}_{\mu}=e^{-\epsilon k((-1)^i s_{\rm{phys}}-r_i)}x_{\mu}\ .
\label{canonictf}
\end{equation}
According to \re{B1eq} and \re{opB1}, we have
\bea
{\widetilde{\square}}_4 \phi_n^{[s]}&=&
e^{2\epsilon k\left((-1)^i s_{\rm{phys}}-r_i\right)}\,\square_4 \phi_n^{[s]}
     \nonumber\\
&=&e^{2\epsilon k\left((-1)^i s_{\rm{phys}}-r_i\right)}\,\left(m_n\right)^2\,\phi_n^{[s]}
    \nonumber\\
&=&\quad \left(m_n^{{\rm{phys}}}\right)^2\,\phi_n^{[s]}\ .
\label{massquare}
\eea
The mass as seen in the brane at $s=s_{\rm{phys}}\in I_i$ is then
\begin{equation}
m_n^{{\rm{phys}}}=e^{\epsilon k\left((-1)^i s_{{\rm{phys}}}-r_i\right)}\,m_n\ .
\label{massquare2}
\end{equation}
For $s_{\rm{phys}}=0$, the physical mass 
is just equal to the mass eigenvalue. 
At the singular point $s_{i+1}$ between $I_i$ and $I_{i+1}$ 
the physical mass is continuous in $s_{\rm{phys}}$.
In the case of a single singularity $s_1$, formula \re{massquare2} becomes
\bea
{\rm{for\ }} 0\leq s_{\rm{phys}} \leq s_1& & m_n^{{\rm{phys}}}
       =e^{\epsilon k s_{{\rm{phys}}}}\,m_n
    \nonumber\\
{\rm{for\ }} s_1\leq s_{\rm{phys}} \leq 2\pi R& & m_n^{{\rm{phys}}}
       =e^{-\epsilon k (s_{{\rm{phys}}}-2s_1)}\,m_n\ .
\label{massquare3}
\eea

As one moves $s_{\rm{phys}}$ away from zero, the physical masses
$m_n^{{\rm{phys}}}$ increase or decrease exponentially. 
The physical masses may therefore differ appreciably
from the eigenvalues. To preserve the mass hierarchy 
solution, the major parameter
$kR$ has to be adequately adjusted. 

%\newpage

\subsection{The Probability densities \label{probdens}}

In the context of a given boundary case,
once all the parameters are fixed and the
mass eigenvalue tower is determined, there exists
a unique field $\phi^{[s]}_n(s)$
for each mass eigenvalue
leading to a naive normalized
probability density field distribution $D_n(s)$
along the fifth dimension \re{scalprodgen}
\begin{equation}
D_n(s)=\frac{\sqrt{g}(\phi^{[s]}_n(s))^2}
        {\int_0^{2\pi R} ds\sqrt{g}(\phi^{[s]}_n(s))^2}\ .
\label{probdensity}
\end{equation}

As discussed at length in \cite{GN2}, 
the probability densities are fast varying functions of $s$.
In a large part of the domain, their logarithms increase
or decrease linearly.

In the brane at $s_{\rm{phys}}$, it is
directly possible to compare the
probability densities of the different mass eigenstates in a given tower.
Neglecting dynamical and
kinematical effects
related to the production in the available phase space, these
probabilities
would account for the rate of appearance
of the mass eigenvalue states to an observer
sitting at this $s_{\rm{phys}}$. Remember however that the physical masses,
as seen by this observer (at $s_{\rm{phys}}\neq 0$), are not the mass eigenvalues
but vary with the $s_{\rm{phys}}$ in agreement with \re{massquare2}.

%\newpage

\subsection{Scalar of non zero mass in the bulk \label{nonzero}}

If instead of
a five-dimensional massless
scalar field, one considers a scalar field of mass $M$,
propagating in the bulk, 
the basic equation \re{dalemb5} becomes
\begin{equation}
\frac{1}{\sqrt{g}}\partial_A \sqrt{g}g^{AB}\partial_B\Phi =-M^2\Phi\ .
\label{dalemb5b}
\end{equation}
In the flat case, 
the square of the Kaluza-Klein mass eigenvalues
are simply shifted by $M^2$ and
become $m_n^2{+}M^2$. 
This is not the case in a warped space 
as the Kaluza-Klein reduction equations \re{B1eq}, \re{B2eq},
even in the case without singularity (or at the left of the first singularity),  
become 
\bea
\square_4\  \phi^{[x]}_n(x^{\mu})
                  &=&-m_n^2\ \phi^{[x]}_n(x^{\mu}) 
    \nonumber\\    
e^{2\epsilon k s}\partial_s e^{-4\epsilon  k s}\partial_s \phi_n^{[s]}(s)
     &=&-\left(m_n^2+M^2 e^{-2\epsilon k s}\right)\phi_n^{[s]}(s) \ .
\label{fivedimmass}
\eea
One sees that the Kaluza-Klein fields and mass eigenvalues
are solutions of different equations. 

%\newpage

\section{Single metric singularity.
Specific boundary conditions and numerical evaluations \label{Onesingularity}}

\subsection{Choice of boundary conditions {\label{illustration}}}

As discussed in Sec.\re{physdis1}, we restrict ourselves 
to boundary conditions corresponding to a single periodic or antiperiodic 
singularity at $s_1$
and to edge point boundary conditions of the form \re{boundB2N} or \re{boundCspe}.
The independent sets of boundary conditions 
that we are using in the numerical evaluations
are summarized in Table\re{tableBCdeux}.
They correspond
to sets obtained in the flat space \cite{GN1} and in the warped space 
when there are no singularities \cite{GN2} 
with an extra $T$ factor
\begin{equation}
T=e^{-4k \epsilon \left(\pi R - s_1\right)} \ .
\label{Tdef1}
\end{equation} 
Each choice of the parameters $\alpha_i, \dots$ 
within a chosen set is a concrete example of boundary conditions.  
Remark that for $s_1=\pi R$ 
the allowed boundary conditions are those of the totally flat case 
for which $T=1$ 
(see \re{BCwarpsing2} and \cite{GN1}).
We showed
in Sec.\re{circle}and Sec.\re{physdis1} that, when $s_1=\pi R$ and thus $T=1$,  
the closure of the strip into a circle with the closure point 
chosen as a periodic or antiperiodic
singularity \re{circleBC}
is possible.
This corresponds to a subcase of the Case A2 of Table\re{tableBCdeux}
($\alpha_3=0,\ \alpha_1=\delta_0$, see also Sec.\re{yone}).

%\newpage

%\newpage

\subsection{Choice of $\overline{k}$ and scaling \label{kscaling}}

Let us remark that the value of $\overline{k}$ can be adjusted arbitrarily by using 
a scale invariance as explained in App.\re{kvalue} and can hence be fixed to
\begin{equation}
\overline{k}=1\ .
\label{kfixed}
\end{equation}

%\newpage

\subsection{Numerical evaluations {\label{numerical}}}

The Kaluza-Klein towers can easily be studied numerically. 
Let us give some illustrative results in the situation 
when there is a single singularity 
located on the strip
$[0,2\pi R]$ at 
\begin{equation}
s_1= y_1\pi R\ ,\quad 0\leq y_1 \leq 2
\label{ydef}
\end{equation}
with the metric \re{onesingular} in which we choose
\begin{equation}
\epsilon =1\ .
\label{epsilon}
\end{equation}

In this section, we concentrate on boundary
conditions belonging to the Case A2 of Table\re{tableBCdeux}. 
This case is particularly interesting as the choice
$\alpha_1=\pm 1$, $\alpha_3=0$ is 
one which allows 
the closure of the strip into a circle \re{circleBC}
(see however the discussion in Sec.\re{circle}).
The other cases 
of boundary conditions (Table \re{tableBCdeux}) follow analogous patterns. 
Some numerical results are given for the Case A5. 

%\newpage

\subsubsection{The limiting case $s_1=2\pi R$ ($ y_1=2$) 
{\label{limitcase}}}

It is obvious that the 
situation of our preceding paper (no singularity) 
\cite{GN2} corresponds here to the limiting case of the strip with 
a single singularity pushed to $s_1=2\pi R$ ($y_1=2$). 
For the same choice of the dimensionless parameter   
\begin{equation}
kR={\overline{k}}{\overline{R}}=6.3\ ,
\label{kR2}
\end{equation}
we checked in a few cases that the resulting 
low lying mass eigenvalues of the Kaluza-Klein towers 
are identical to those 
evaluated according to our preceding article.
Considering that the low lying mass eigenvalues are 
of the order of one TeV and that they are equal to the   
physical masses for a four-dimensional observer at $s_{\rm{phys}}=0$ (see \re{massquare2}),
the hierarchy problem is seen to be solved in the sense that imposing a unique mass scale 
($m_{\rm{Pl}}$), the TeV mass scale is recovered. See in particular the first 
line of Table \re{tableKK63} which corresponds to the Case A2 of Table\re{tableBCdeux} 
with the choice of parameters
\begin{equation} 
\alpha_1=1,\ \alpha_3=0,\ \overline{k}=1,\ kR=6.3,\ \delta_1=1,\ \epsilon=1\ .
\label{a2}
\end{equation}

%\newpage

\subsubsection{Case A2. Arbitrary location of the singularity at $s_1=y_1\pi R$ with $0\leq y_1\leq 2$
{\label{arbitrary}}}

In the here above Case A2 \re{a2},
we have studied the consequences of the presence of the periodic singularity 
(see \re{transition}, \re{periodicdef})
when $y_1$ is decreased 
from $2$ down to $0$.

The resulting low lying mass eigenvalues for fixed $kR=6.3$ 
are listed in Table \re{tableKK63} . 
As we already said, in the limiting case  $y_1{=}2$, the situation 
is as if there was no singularity. 

One sees that, when $y_1$ decreases, the mass eigenvalue towers have a
small mass $m_1$ 
which decreases slightly and levels to $0.16$ TeV. 
The higher order masses $m_2,m_3,\dots$ increase drastically, spoiling 
badly the mass hierarchy solution already for $y_1\approx 1.7$.
It can be restored by increasing $kR$ progressively,
for example to $kR\approx 8.4$ for $y_1=1.5$ as can be seen in Table \re{tableKK15}.
It appears that, as a general rule, the mass hierarchy solution can be restored 
for all values of $y_1$
by adopting for $kR$ the approximate value 
\begin{equation}
kR\approx \frac{12.6}{y_1}
\label{approx}
\end{equation} 
as can be seen in Table \re{tableKK126}.
The mass eigenvalues $m_2,m_3,\dots$ are decreasing slowly for decreasing
$y_1$ and stabilize already from $y_1\approx 1.8$ downwards.  
The mass eigenvalue $m_1$ has a peculiar behavior, 
decreasing sharply to zero for 
$y_1\rightarrow 1$ and then increasing to a small limiting value 
which is already reached at $y_1=0.9$. 
We have decided not to include values for $y_1$ smaller than $0.05$ in the Table 
as $kR$ \re{approx} then violates the one-mass-scale-only postulate \re{barpara}.

Still considering the Case A2 \re{a2} but relaxing the restriction $\alpha_3=0$, 
the zero mass constraint 
(see Table \re{tablemasszero}) leads to a curve 
$\alpha_3^{[0]}$ as a function of $y_1$. This curve is always above the
$y_1$ axis and is 
tangent to it at $y_1=1$. For 
$\alpha_3<\alpha_3^{[0]}$,
a particle appears at the bottom of the Kaluza-Klein tower; 
for $\alpha_3>\alpha_3^{[0]}$,
a tachyon is present (see Table \re{tableKK195}).  

%\newpage

\subsubsection{Case A2. Location of the singularity at $s_1=\pi R$ ($y_1=1$) {\label{yone}}. 
Closure into a circle }

In the case A2 \re{a2} with $y_1=1$ and $\alpha_3=0$, 
one reaches the situation where the strip can be
closed into a circle, with a second singularity at 
$\{s=2\pi R\}\equiv \{s=0\}$ and $\alpha_1$ being identified with $\delta_0$. 
Both singularities can be taken independently 
as periodic or antiperiodic $\delta_0=\pm 1,\ \delta_1=\pm 1$.
If both have the same periodicity $\alpha_1\delta_1=1$,
the zero mass condition 
(see Table \re{tablemasszero})
is exactly satisfied. 
This agrees with line $y_1=1$ in Table \re{tableKK126}.

%\newpage

\subsubsection{Case A2. Closing into a circle.
Some points of comparison with the original Randall-Sundrum scenario
{\label{comparison}}}

$\ \bullet$
Rizzo \cite{Rizzo} has elaborated on the Kaluza-Klein towers
in the Randall-Sundrum scenario. He stated that the masses
are related to the roots $z_p$ of the first Bessel function $J_1(z)$ 
by 
\begin{equation}
m_p=e^{\pi k R}\frac{z_p}{k}
\label{Rizzomass}
\end{equation}
and would be interpreted as the physical masses in the TeV brane 
which he takes at $s_1=\pi R$.
The mass sequence as illustrated on his Figure 5 corresponds 
to the choice of $k R\approx 11$ and $\overline{k}=k/M_{\rm{Pl}}\approx 0.01$.
These masses
are listed in the first line of Table \re{RizzoGN}.

The situation considered by Rizzo is equivalent to our Case A2 of Table\re{tableBCdeux} 
($\alpha_3{=}0$, $\epsilon{=}\alpha_1=\delta_0=\delta_1=1$) with the strip closed into a circle 
with two periodic singularities located at $y_1=0$ and $y_1=1$. Adopting the same parameters 
$k R=11$ and $\overline{k}=0.01$, we obtain the mass eigenvalues 
which are listed in the second line of Table \re{RizzoGN}.

By inspection of this Table, one sees that the even indexed masses 
$m_{2n}$ agree.
Obviously, one mass out of two is absent in the mass tower 
as established by Rizzo, but this is due to the $Z_2$ symmetry $s\rightarrow -s$. 
Our states are even or odd under
$Z_2$ while  Rizzo selected the even states for orbifold reasons.  

\noindent$\ \bullet$
A basic ingredient of the Randall-Sundrum scenario is the existence of the so-called
visible brane which is located at
$s=0$ (see \cite{RS2}
correcting \cite{RS1}). With all the parameters in the bulk 
scaled with the Planck mass, 
the corresponding low lying physical masses in this visible 
brane are of the order of the TeV.

Here we would like to stress that our warp model, 
with or without metric singularities,
on a strip or on a circle, 
although inspired by the Randall-Sundrum approach, 
has been developed independently in a
mathematically consistent and complete way
(up to dynamical considerations). 
The only mass scale is also the Planck mass. 
The warped parameters $k$ and $R$ are given values such that the 
low lying Kaluza-Klein 
eigenvalues are of the order one TeV,
thus solving the hierarchy problem. 
A typical aspect of our model resides 
in the fact that the physical Kaluza-Klein 
masses as measured by a four-dimensional observer 
are deduced from these eigenvalues by formula \re{massquare2}.
Hence they depend 
on the location of the 
physical brane, which can be anywhere on the 
extra dimension axis.   

%\newpage

\subsubsection{Case A5
 {\label{casea5}}}
 
The numerical results in the Case A5 
are summarized in the Tables \re{caseA5a} and \re{caseA5b}
for $\overline{k}=1$ and with $kR=12.6/y_1$. 
The Kaluza-Klein eigenvalue spectrum is composed of two different 
components with different behaviors.

\noindent$\ \bullet$
One component (Table \re{caseA5a}) consists in
a tower of masses which depend on $y_1$ and not
on $\overline{\zeta}$ in a very large range of $\overline{\zeta}$ 
including the natural range \re{barpara}.
Decreasing $y_1$ from $y_1{=}2$, the situation with no singularity, the 
low lying eigenvalues 
converge to the same limiting spectrum already for $y_1{=}1.9$.
 
\noindent$\ \bullet$
The second component consists in a
lonely eigenvalue which is essentially independent of $y_1$ but depends
steeply on $\overline{\zeta}$. 
The eigenvalue is zero for $\overline{\zeta}^{[0]}=4 E^{2(1-y_1)}/F$ 
(see line A5 of \re{tablemasszero}).
This $\overline{\zeta}^{[0]}$ turns out to be 
weakly dependent on $y_1$. It 
is negative and lies in the restricted range
$-6.89\ 10^{-69}< \overline{\zeta}^{[0]} < 3.43\ 10^{-69}$.  
For $\overline{\zeta}>\overline{\zeta}^{[0]}$ the eigenvalue
corresponds to a particle state. For $\overline{\zeta}<\overline{\zeta}^{[0]}$
it corresponds to a tachyon.

In the whole range
\begin{equation}
-10^{-31}\leq \overline{\zeta}\leq 10^{-31}\ , 
\label{A5range}
\end{equation}
the eigenvalue is low lying and varies
essentially as
\begin{equation}
m^2\approx 2\left(\zeta-\zeta^{[0]}\right)k\ .
\label{A5masseq2}
\end{equation}
The eigenvalues are listed in Table \re{caseA5b}.

It should be remarked that this 
second component violates the one-mass-scale-only hypothesis
as $\overline{\zeta}$ must be fine tuned \re{A5range} to a very small number.

\noindent$\ \bullet$
Both results above can be understood 
from the expression of the determinant 
(for $m^2$ positive)
which provides the eigenvalues. 
Its leading term is the product of two factors. 
One factor is independent of $\overline{\zeta}$ and 
its roots provide the tower as the first component.
The second factor is independent of $y_1$ 
\begin{equation}
\zeta\, Y(2,\frac{m}{k})-m\, Y(1,\frac{m}{k})  
\label{A5masseq1}
\end{equation}
and its root give the second component 
in agreement with Eq.\re{A5masseq1} when $m$ is low lying and hence
$m/k$ is small.
The proof for the tachyon case is analogous.

%\newpage

\section{Conclusions {\label{conclusion}}}

In this article, we have extended the warp model that we developed in
our previous paper \cite{GN2} by the inclusion of one or more singularities
in the metric.
The metric we adopted is  related to a five-dimensional warped
space arising from a constant negative bulk cosmological constant, with the
fifth extra dimension being compactified either on a strip or in some cases
on a circle. The metric singularities are located at some fixed points in
the extra dimension range where continuity conditions are imposed to the
metric.
We showed in particular that the strip can be closed into a
circle when the number of singularities is even and when the total range in
the  extra dimension where the metric is increasing is equal to 
the total range where it is decreasing.

We considered again a five-dimensional massless real scalar field supposed to
propagate in the bulk and followed closely the discussion in \cite{GN2} 
relative to the hermiticity and commutavity properties of the
operators entering in the Kaluza-Klein reduction equations, 
thereby ensuring the existence and reality of the mass eigenvalues.
Taking into account the presence 
of metric singularities, we generalized 
all the allowed sets of boundary conditions to be imposed on
the field solutions of the Kaluza-Klein reduction
equations.
For each set of boundary conditions 
and for some choice of the parameters fixing
them, one can deduce either the mass eigenvalues 
building up a so called Kaluza-Klein mass
tower, or eigenvalues related to a tachyon. 
For each set, there is 
a surface 
in parameter space where one mass eigenvalue
is zero, with on one side mass states and on the
other side tachyon states. Close to the surface, 
the masses squared, positive or negative, are small.

Our basic assumption is that there is one-mass-scale-only in
our model, namely the Planck scale. By a choice, unique for all
boundary conditions, compatible with this assumption, of the two
major parameters of the model, 
$k$ the warp factor and $R$ measuring the
extension of the extra dimension, one 
solves the mass hierarchy problem,
in the sense that the resulting low lying
mass eigenvalues are of the order of one TeV.

A specific aspect of our model resides in the fact that in a
brane, the brane of a four-dimensional observer, a Kaluza-Klein 
eigenvalue tower
appears as a tower of physical masses which are equal to the eigenvalues
multiplied by a factor depending on the position of the brane in the 
extra dimension. The coordinate of this position 
is an arbitrary parameter of the
model.

Finally, we have illustrated our theoretical results by some
numerical evaluations in a few 
boundary condition cases with 
a single metric singularity. 
Moving the singularity along the extra dimension axis generally 
results in very large variations of the mass eigenvalues. In order to save 
the mass hierarchy
solution, it appears that the dimensionless parameter $kR$ has to be given values inversely
proportional to the coordinate of the singularity.
It should be noticed that the case where the strip can be closed into a circle, 
with two singularities at $0$ and $\pi R$,
gives a Kaluza-Klein tower which is practically the same as those 
corresponding to a single singularity located anywhere in a 
wide range around $\pi R$,
except that it has a zero mass state. 

\newpage

\appendix

\section{General discussion of the Principle of ``Least Action''  \label{least}}

In this appendix, we give a detailed 
and general discussion of the path leading to the
derivation of the equations of motion of a free massless complex scalar field 
in a warped five-dimensional 
space with the fifth dimension compactified and with a single metric singularity. 
The extension to more metric singularities or to a massive field is straightforward.

The invariant 
scalar product is \re{scalprodgen}. 
The corresponding most general invariant action is quadratic in the field
\bea
{\cal{A}}
&=&\int_{-\infty}^{+\infty} d^4x\, \int_0^{2\pi R}ds
\ \Biggl\{a\,(\partial_A\Phi^*)\,\sqrt{g}\,g^{AB}\,(\partial_B \Phi)\Biggr.
      \nonumber\\
       &&\quad\quad\quad\quad\quad\quad\quad\quad\quad\quad\quad\quad
       +\,b\ \Phi^*\,\partial_A\biggl(\sqrt{g}\,g^{AB}(\partial_B\Phi)\biggr)
      \nonumber\\
       &&\Biggl.\quad\quad\quad\quad\quad\quad\quad\quad\quad\quad\quad 
       +\,c\ \partial_A\biggl((\partial_B\Phi^*)\sqrt{g}\,g^{AB}\biggr)\,\Phi
        \Biggr\}\ .
\label{action1sing}
\eea
Let us make a few comments
\begin{enumerate}

\item

The Lagrangian is Hermitian for
\begin{equation}
a\ {\rm{real}}\quad,\quad c=b^*\ .
\label{reality}
\end{equation}

\item 

Since we postulate a singularity at $s=s_1$, one has to split 
the integration domain in $s$ into two regions $[0,s_1]$ and $[s_1,2\pi R]$ 
and study carefully what happens at the four 
boundary points.  
Besides the end points $0$ and $2\pi R$, we anticipate that 
the values of
$\Phi$ and of its $s$-derivative on the left $s_1{-}\eta$  and on the right 
$s_1{+}\eta$ of the singularity ($\eta\rightarrow 0^+$) 
play a role in the boundary conditions.

\item

It is well-known that the three parts 
(with coefficients $a,b,c$)
of the action \re{action1sing} lead to the same Euler-Lagrange equation. 
Indeed, they differ in the integrand 
by total derivatives, hence by boundary terms only. The differences depend 
on the values of the fields and of their derivatives at 
all the edges of the $s$ range. 
When the action is varied (in view of finding solutions according to 
the ``least action principle''), and
variations of the fields at the edges are taken into account, 
the three parts of the Lagrangian are not equivalent, as we will now discuss.

\item

For the four space-time integrations, on $x^{\mu}$, the fields (belonging to the Hilbert space) 
must decrease sufficiently fast at $x^{\mu}\rightarrow \pm \infty$, so that the boundary values
of the variations of the fields do not play any role. 
The finite range of the extra dimension $s$ requires a more careful 
treatment.  

\item
According to most textbooks, the Euler Lagrange equations are obtained by requesting 
the variation of the action to be zero for arbitrary variations of the fields
keeping them
zero at the boundaries. Here, we suppose, in the variable $s$, that
\bea
(\delta\Phi)(0)&=&(\delta\Phi)(2\pi R)= 0
        \label{varzero1}\\
\left(\delta\left(\partial_s\Phi\right)\right)(0)&=&\left(\delta\left(\partial_s\Phi\right)\right)(2\pi R)= 0
\label{varzero2}
\eea
and that the fields and their variations are continuous at the singularity point $s_1$
($\eta\rightarrow 0^+$)
\bea
\Phi(s_1-\eta)&=&\Phi(s_1+\eta)
        \label{varzero3}\\
(\partial_s\Phi)(s_1-\eta)&=&(\partial_s\Phi)(s_1+\eta)
        \label{varzero4}\\
(\delta\Phi)(s_1-\eta)&=&(\delta\Phi)(s_1+\eta)
        \label{varzero5}\\
(\delta(\partial_s\Phi))(s_1-\eta)&=&(\delta(\partial_s\Phi))(s_1+\eta)\ .
\label{varzero6}
\eea
One finds
\bea
\delta{\cal{A}}
&=&\Biggl(-a+b+b^*\Biggr)
      \int_{-\infty}^{+\infty} d^4 x\left(\int_{0}^{s_1}+\int_{s_1}^{2\pi R}\right)ds
      \nonumber\\
      &&\!\!\!\!\!\!\!\!\!\!\!\!\Biggl((\delta\Phi)^*\partial_A\biggl(\sqrt{g}\,g^{AB}(\partial_B\Phi)\biggr)
      +\partial_A\biggl( \sqrt{g}\,g^{AB}(\partial_B\Phi^*\biggr)\left(\delta\Phi\right)
      \Biggr)      
     \nonumber\\
     &\equiv&\delta{\cal{A}}_{\rm{core}}\ .
\label{deltacore}
\eea
For the term with coefficient $a$ 
in \re{action1sing}, it suffices to impose \re{varzero1}, \re{varzero4} and \re{varzero5} 
while for the terms with $b$ and $b^*$, one needs all the conditions from \re{varzero1} to \re{varzero6}.
Usually, the term $a$ only is taken into account. 
The vanishing of $\delta{\cal{A}}$ under
arbitrary variations of $\delta\Phi$ then leads to the Riemann equation \re{dalemb5} under the lone condition
\begin{equation}
a-b-b^* \neq 0\ .
\label{abcnorm}
\end{equation}

\item

Let us analyze the problem when no restrictions at all are imposed 
a priori at the boundaries, 
i.e. none of \re{varzero1}-\re{varzero6}. The variation $\delta{\cal{A}}$ 
can then be decomposed into two terms
\begin{equation}
\delta{\cal{A}}=\delta{\cal{A}}_{\rm{core}}+\delta{\cal{A}}_{\rm{bound}}
\label{deltaA}
\end{equation}
with
\bea
\delta{\cal{A}}_{\rm{bound}}&=&\int_{-\infty}^{+\infty} d^4 x\left(\int_{0}^{s_1}+\int_{s_1}^{2\pi R}\right)ds
            \nonumber\\
      &&\partial_A\Biggl[\ b\ \Phi^*\sqrt{g}\,g^{AB}\bigl(\partial_B\left(\delta\Phi\right)
                   \bigr)\Biggr.
           \nonumber\\
      &&\phantom{\partial_A}+\left(a-b\right)\left(\partial_B\Phi^*\right)\sqrt{g}\,g^{AB}\left(\delta\Phi\right)
           \nonumber\\
     &&\phantom{\partial_A}+\ b^*\ \left(\partial_B(\delta\Phi^*)\right)\,\sqrt{g}\,g^{AB}\ \Phi
           \nonumber\\
     &&\phantom{\partial_A}+\left(a-b^*\right)\Biggl.\ \left(\delta\Phi^*
     \right)\,\sqrt{g}\,g^{AB} \left(\partial_B\Phi\right)
                   \Biggr]\ .
\label{deltabound}
\eea
We define
\bea
{\cal{B}}^A(s)&=&\int_{-\infty}^{+\infty} d^4 x
    \Biggl[\phantom{+}
\ b\ \Phi^*\sqrt{g}\,g^{AB}\bigl(\partial_B\left(\delta\Phi\right)
                   \bigr)\Biggr.
           \nonumber\\
      &&\phantom{mmmmmm}+\left(a{-}b\right)\left(\partial_B\Phi^*\right)\sqrt{g}\,g^{AB}\left(\delta\Phi\right)
           \nonumber\\
     &&\phantom{mmmmmm}+\ b^*\ \left(\partial_B(\delta\Phi^*)\right)\,\sqrt{g}\,g^{AB}\ \Phi
           \nonumber\\
     &&\phantom{mmmmmm}+\left(a{-}b^*\right)\Biggl.\ \left(\delta\Phi^*\right)\,\sqrt{g}\,g^{AB} \left(\partial_B\Phi\right)\Biggr]
\ .
\label{defB}     
\eea
The only term which contributes 
to $\delta{\cal{A}}_{\rm{bound}}$, for the metric \re{metric}, is for the indices $A=B=5\equiv s$ with $g^{55}=-1$. It leads to 
the necessary action boundary relation for the fields and their variations 
\begin{equation}
\delta{\cal{A}}_{\rm{bound}}=\lim_{\eta\rightarrow 0^+}
%\int_{-\infty}^{+\infty} d^4 x
      \Biggl[
       {\cal{B}}^s(2\pi R) -{\cal{B}}^s(s_1{+}\eta)+{\cal{B}}^s(s_1{-}\eta)-{\cal{B}}^s(0)                      
                      \Biggr]=0
\label{deltaAbound}
\end{equation}
from which the action boundary conditions have to be determined.

\item

We perform the Kaluza-Klein reduction \re{KKreduc} on $\Phi(x,t)$ and denote $\Phi$ by \re{appnotation}
in terms of the Kaluza-Klein reduced fields $\phi_n^{[s]}$. 
The Kaluza-Klein reduced fields 
$\phi^{[s]}_n(s)$ belong to a Hilbert space defined by boundary conditions of the
form \re{BCN} with a $4\times 8$ matrix $M$ of rank four. 
It is reasonnnable to suppose that the field variations $\delta\phi^{[s]}_p(s)$ 
belong to the same Hilbert space. In other word, the action is varied within that Hilbert space.
We denote by $\Theta$ the vector analogous to $\Phi$ \re{vectorphi} built out from
$\delta\phi^{[s]}_p(s)$ and its derivative. Namely 
\bea
\phantom{\Biggl\{\Biggr\}}
      \theta_0&=&(\delta\phi^{[s]}_p)(0)
      \nonumber\\
\phantom{\Biggl\{\Biggr\}}
      \partial\theta_0&=&(\partial_s(\delta\phi^{[s]}_p))(0)
      \nonumber\\      
\phantom{\Biggl\{\Biggr\}}
      \theta_l&=&e^{-2\epsilon k s_1}\lim_{\eta\rightarrow 0^+}(\delta\phi^{[s]}_p)(s_1-\eta)
      \nonumber\\
\phantom{\Biggl\{\Biggr\}}
      \partial\theta_l&=&e^{-2\epsilon k s_1}\lim_{\eta\rightarrow 0^+}(\partial_s(\delta\phi^{[s]}_p))(s_1-\eta)
      \nonumber\\      
\phantom{\Biggl\{\Biggr\}}
      \theta_r&=&e^{-2\epsilon k s_1}\lim_{\eta\rightarrow 0^+}(\delta\phi^{[s]}_p)(s_1+\eta)
      \nonumber\\
\phantom{\Biggl\{\Biggr\}}
      \partial\theta_r&=&e^{-2\epsilon k s_1}\lim_{\eta\rightarrow 0^+}(\partial_s(\delta\phi^{[s]}_p))(s_1+\eta)
      \nonumber\\      
\phantom{\Biggl\{\Biggr\}}
      \theta_R&=&e^{4\epsilon (\pi R -s_1)}(\delta\phi^{[s]}_p(R)
      \nonumber\\
\phantom{\Biggl\{\Biggr\}}
      \partial\theta_R&=&e^{4\epsilon (\pi R -s_1)}(\partial_s(\delta\phi^{[p]}_p))(0) \ .
\label{appdelta}     
\eea    

\item

With this notation, the action boundary relation \re{deltaAbound} (after the Kaluza-Klein reduction) is written
\begin{equation}
\Theta^+T^{[8]}\Phi + \Phi^+T^{[8]+}\Theta=0
\label{BCwarpsingaction}
\end{equation}
where $T^{[8]}$ is the $8\times 8$ matrix 
\begin{equation}
T^{[8]}=b^*\,S^{[8]}+a\,U^{[8]}
\label{matrixT8}
\end{equation}
with the matrices $S^{[8]}$ \re{matrixS8} and 
\begin{equation}
U^{[8]}=\left(
     \matrix{
           U^{[4]} & 0^{[4]}\cr
           0^{[4]} & U^{[4]}
           }=1^{[4]}\otimes U^{[4]}
\right)
\label{matrixU8}
\end{equation}
\begin{equation}
U^{[4]}=\left(
     \matrix{
           \sigma_+ & 0^{[2]}\cr
           0^{[2]} & -\sigma_+ 
           }=\sigma_3\otimes \sigma_+
\right)
\label{matrixU4}
\end{equation}
with $\sigma_+=(\sigma_1{+}i\sigma_2)/2$.

Following the same procedure as in Sec.\re{HBC}, one obtains for  $\Phi$ and $\Theta$
boundary relations and conditions similar to those obtained 
for $\Phi$ and $\Psi$ from the hermiticity of the Riemann operator
\re{boundrel} with $S^{[8]}$ replaced by $T^{[8]}$.

\item

For the common Hilbert space that we hypothized for $\delta\Phi$ and $\Phi$ to be the 
exactly the same that
we obtained in Sec.\re{HBC},
we are led to impose the (unusual) restriction
\begin{equation}
 a=0
\label{zeroa}
\end{equation}
in the formulation of the initial Lagrangian \re{action1sing}.

\item 

In summary, we find that in order
to obtain our sets of allowed Hilbert spaces 
we can adopt two different options leading to exactly the same consequences
in terms of sets of allowed boundary conditions.

\begin{description}
	
\item{\bf{Option 1}} 

Start with any action of the form \re{action1sing}, \re{reality} 
with the restriction \re{abcnorm} and apply the ``least action principle'' 
with vanishing variations at the edges of $s$ \re{varzero1}-\re{varzero2} 
and continuity at the singular point \re{varzero3}-\re{varzero3} to obtain
the Riemann operator \re{dalemb5}. 
The requirement that this operator be self-adjoint (after Kaluza-Klein reduction)
leads to the boundary conditions for the fields and hence 
to the allowed relevant Hilbert spaces.

\item{\bf{Option 2}} 

Start with an action of the form \re{action1sing}, \re{reality} 
with the restriction $a=0$ \re{zeroa} and apply the ``least action principle'' 
with the fields $\Phi$ and their variation $\delta\Phi$ belonging to the same Hilbert space.
This in fact leads  to boundary conditions identical to those of Option 1. On this Hilbert space, 
the Riemann operator turns out to be automatically selfadjoint.

\end{description}

\end{enumerate}

%\newpage

\section{Examples of allowed boundary conditions \label{appex}}

In this appendix, we give a few examples of boundary conditions 
which derive from the general considerations given in Sec.\re{HBC}
for some choices of the permutation $P$.

\begin{description}

\item{Case A}

Suppose first that $P = A\equiv 1^{[8]}$. Hence $\Phi_A=\Phi$ \re{vectorphi},
\begin{equation}
\Phi_A^u=\left(
     \matrix{
           \phi_0\cr
           \partial\phi_0\cr
            \phi_l\cr
           \partial\phi_l\cr
           }
\right)
\label{vectorphiAu}
\end{equation}
and
\begin{equation}
\Phi_A^d=\left(
     \matrix{
           \phi_r\cr
           \partial\phi_r\cr
            \phi_R\cr
           \partial\phi_R\cr
           }
\right)\ ,
\label{vectorphiAd}
\end{equation}
and the boundary conditions \re{BCcaseP} are written
\begin{equation}
\Phi_A^d=V_A^{[4]}\Phi_A^u
\label{BCcaseA}
\end{equation}
which means that, in this case, the four field (and derivative) 
boundary values at the right of the singularity (at $r$ and $R$) are
linear functions of the four boundary values on the left (at $0$ and $l$). 

With the matrix $S^{[8]}_A=S^{[8]}$ \re{SP1},\re{matrixS8} 
and the form \re{matrixV8} for $V_P^{[8]}$, 
the equation for $V_A^{[4]}$  originating from \re{SPBC},\re{BCP} is
\begin{equation}
S^{[4]} = -V_A^{[4]+}\,S^{4}\,V^{[4]}_A\ .
\label{condcaseA}
\end{equation}
Defining 
\begin{equation}
Q^{[4]}=\left(
     \matrix{
           0^{[2]}&1^{[2]}\cr
           1^{[2]}&0^{[2]}
           }
\right)\ ,
\label{matrixQC4}
\end{equation}
the matrix $W^{[4]}_A = Q^{[4]}V^{[4]}_A$ satisfies $S^{[4]}_A = W^{[4]+}_A S^{[4]}_A W^{[4]}_A$ and hence is
complex-symplectic. 
Inversely $V_A^{[4]}$
must be a complex-symplectic matrix multiplied on the left by $Q^{[4]}$.
Consequently $\mid \det V_A^{[4]} \mid= 1$ and $V_A^{[4]}$ is
invertible. Let us recall that
the space of complex-symplectic matrices $W_A^{[4]}$ depends on 16 arbitrary real parameters. 
In the case of a
real scalar field, $W^{[4]}_A$ must be real-symplectic
and there are 10 real parameters.

Any specific choice of the 16 real parameters 
(or 10 for the real fields) 
leads to an allowed set of boundary conditions.

Since $\det V^{[4]}\neq 0$, the interchange of $\Phi_A^u$ and $\Phi_A^d$ by
\begin{equation}
P=\left(
     \matrix{
           0^{[4]} & 1^{[4]}\cr
           1^{[4]}&0^{[4]}
           }
\right)
\label{detA}
\end{equation}
leads to equivalent boundary conditions.

There is one important subcase of Case A worth mentioning.

\item{Case B}

The preceding case ($P=1$) with $V_A^{[4]}$ restricted to be of the form
\begin{equation}
V_{B}^{[4]}=\left(
     \matrix{
           0^{[2]} & V_{B2}^{[2]}\cr
           V_{B3}^{[2]}&0^{[2]}
           }
\right)
\label{detBCB}
\end{equation}
is physically interesting.
The fields (and their derivatives) evaluted at $s = 2\pi R$
are linearly related to those evaluated at $0$. Those at both sides of
the singularity ($s = s1 - \eta$ and $s = s1 + \eta$) are related by other linear
relations. 
The matrices $V_{B2}^{[2]}$ and $V_{B3}^{[2]}$ must both be complex-sympletic.
In the complex case, it implies that these matrices
are equal to an arbitrary phase factor $e^{i\varphi_{v_j}},\ j=2,3$ multiplied by a real matrix of
determinant one. Collecting the results, this set of boundary
conditions is written
\bea
\left(
     \matrix{
           \phi_r\cr
           \partial \phi_r
           } \right)
         &=&
      e^{i\varphi_{v_2}}
     \left(
     \matrix{
           \alpha&\beta\cr
           \gamma&\delta
           } \right)
           \left(
     \matrix{
           \phi_l\cr
           \partial \phi_l
           } \right)      
     \ \ \,,\quad \alpha\delta-\beta\gamma =1
         \label{boundB1}\\           
\left(
     \matrix{
           \phi_R\cr
           \partial \phi_R
           } \right)
         &=&
      e^{i\varphi_{v_3}}
     \left(
     \matrix{
           \alpha'&\beta'\cr
           \gamma'&\delta'
           } \right)
           \left(
     \matrix{
           \phi_0\cr
           \partial \phi_0
           } \right)      
     \ ,\quad \alpha'\delta'-\beta'\gamma' =1\ .
\label{boundB2}
\eea     
For real fields, we emphasize the particular set
when $\alpha = \delta = 1,\ \beta=\gamma=0$ 
and $e^{i\varphi_{v_2}} = \delta_1$. Namely, at $s_1$, the boundary conditions
are
\bea
\phi_r
           &=&\delta_1\phi_l
      \nonumber\\
\partial \phi_r
          &=&\delta_1\partial \phi_l
\label{transition}
\eea
where $\delta_1$ is an arbitrary sign. We adopt the following denomination convention
for a singularity $s_1$ with that type of boundary conditions
\bea
\delta_1={+}1&\rightarrow &{\rm{periodic\ singularity}}
     \nonumber\\
\delta_1={-}1&\rightarrow &{\rm{antiperiodic\ singularity}}\ .
     \label{periodicdef}
\eea

\item{Case C}     
 
Suppose that the boundary conditions relate the derivatives of the fields
to the fields themselves. This is achieved by the permutation     
\begin{equation}
P_C=\left(
     \matrix{
             1& 0& 0& 0& 0& 0& 0& 0\cr
             0& 0& 1& 0& 0& 0& 0& 0\cr
             0& 0& 0& 0& 1& 0& 0& 0\cr
             0& 0& 0& 0& 0& 0& 1& 0\cr
             0& 1& 0& 0& 0& 0& 0& 0\cr
             0& 0& 0& 1& 0& 0& 0& 0\cr
             0& 0& 0& 0& 0& 1& 0& 0\cr
             0& 0& 0& 0& 0& 0& 0& 1\cr       
           }
\right)\ .
\label{PC}
\end{equation}
This choice of $P_C$ leads to
\begin{equation}
\Phi_C^u=\left(
     \matrix{
           \phi_0\cr
           \phi_l\cr
           \phi_r\cr
           \phi_R\cr
           }
\right)
\label{vectorphiCu}
\end{equation}
and 
\begin{equation}
\Phi_C^d=\left(
     \matrix{
           \partial\phi_0\cr
           \partial\phi_l\cr
           \partial\phi_r\cr
           \partial\phi_R\cr
           }
\right)\ ,
\label{vectorphiCd}
\end{equation}
so, the boundary conditions \re{BCV}, \re{matrixV8}, \re{BCcaseP} read
\begin{equation}
\Phi_C^d=V_C^{[4]}\Phi_C^u\ .
\label{BCcaseC}
\end{equation}
The matrix $S_C^{[8]}$ \re{SP1}, \re{SP2} has elements
\bea
S_{C1}^{[4]}=\ \ S_{C4}^{[4]} &=& 0^{[4]}
     \nonumber\\
S_{C2}^{[4]}= - S_{C3}^{[4]} &=& \left(
                     \matrix{\sigma_3& 0^{[2]}\cr
                             0^{[2]}& \sigma_3
                             }
                     \right)     
\label{SB}
\eea
with, from \re{BCP}, \re{SPBC},
the restriction 
for $V_C^{[4]}$
\begin{equation}
V_{C}^{[4]+}\ S_{C2}^{[4]}=S_{C2}^{[4]}\ V_C^{[4]}\ \left(=(S_{C2}^{[4]}\ V_C^{[4]})^+\right)
\ .
\label{BCB}
\end{equation}
In other words, $V_C^{[4]}$ must be equal to an Hermitian matrix multiplied
on the left by $S_{C2}^{[4]}$.
For $V^{[4]}_C$, written in $2\times 2$ block form, one has
\bea
\left(\sigma_3V_{C1}^{[2]}\right)^+&=&\sigma_3 V_{C1}^{[2]}
     \nonumber\\
\left(\sigma_3V_{C4}^{[2]}\right)^+&=&\sigma_3V_{C4}^{[2]}
    \nonumber\\
\left(\sigma_3V_{C3}^{[2]}\right)^+&=&\sigma_3V_{C2}^{[2]}\ .
\label{caseBVB}
\eea
The boundary condition is thus seen to depend also on 16 arbitrary real parameters
for a complex field and on 10 real parameters for a real field.
Remark that in this case $V_C^{[4]}$ is not always invertible.

A particular case is when $V^{[4]}_C$ is diagonal which means Sturm Liouville type 
boundary conditions. The value of the derivative is related 
to the value of the field evaluated at each of the end points.

A case very analogous to Case C is obtained when the fields and their
derivatives are interchanged, which means that the boundary values of
the fields are expressed in terms of the boundary values of the derivatives.
The result is identical {\it{mutatis mutandis}} but often not equivalent.

\item{Case D}

An interesting subcase of Case C is obtained when $V^{[4]}_C$ is diagonal (arbitray
real diagonal elements). This corresponds to Case A3 in Table\re{tableBCdeux}
for the behavior of the fields at the edges $0,2\pi R$ of the $s$-strip 
and to analogous
conditions at the two sides of the singularity.

\item{Case E}

Another peculiar possibility is when the boundary values of the fields
and their derivatives at the singularity are expressed in terms of the
values of the fields and their derivatives at the edges, namely
\begin{equation}
\Phi_E^u=\left(
     \matrix{
           \phi_0\cr
           \partial\phi_0\cr
            \phi_R\cr
           \partial\phi_R\cr
           }
\right)
\label{vectorphiEu}
\end{equation}
and 
\begin{equation}
\Phi_E^d=\left(
     \matrix{
           \phi_l\cr
           \partial\phi_l\cr
            \phi_r\cr
           \partial\phi_r\cr
           }
\right)\ .
\label{vectorphiEd}
\end{equation}
This is achieved with the permutation
\begin{equation}
P_E=\left(
     \matrix{
             1& 0& 0& 0& 0& 0& 0& 0\cr
             0& 1& 0& 0& 0& 0& 0& 0\cr
             0& 0& 0& 0& 0& 0& 1& 0\cr
             0& 0& 0& 0& 0& 0& 0& 1\cr
             0& 0& 1& 0& 0& 0& 0& 0\cr
             0& 0& 0& 1& 0& 0& 0& 0\cr
             0& 0& 0& 0& 1& 0& 0& 0\cr
             0& 0& 0& 0& 0& 1& 0& 0\cr       
           }
\right)\ .
\label{PE}
\end{equation}
The boundary conditions \re{BCV} are written
\begin{equation}
\Phi_E^d=V_E^{[4]}\Phi_E^u\ .
\label{BCcaseE}
\end{equation}
The $8 \times 8$ antisymmetric matrix $S^{[8]}_E$ becomes
\begin{equation}
S_{E}^{[8]}=\left(
     \matrix{
           S_{C}^{[4]}&0^{[4]}\cr
           0^{[4]}&-S_{C}^{[4]}
           }=\sigma_3\otimes S_{C}^{[4]}
\right)
\label{matrixSE8}
\end{equation}
with
\begin{equation}
S_{E}^{[4]}=\left(
     \matrix{
           i\sigma_2&0^{[2]}\cr
           0^{[2]}&-i\sigma_2
           }
\right)=\sigma_3\otimes (i\sigma_2)\ .
\label{matrixSE4}
\end{equation}
The condition on $V^{[4]}_E$ \re{SPBC} is
\begin{equation}
S^{[4]}_E = V_E^{[4]+}\,S^{[4]}_E\,V^{[4]}_E
\label{condcaseE}
\end{equation}
implying that it is a complex-symplectic matrix.
Here, $V^{[4]}_E$ is invertible and hence a set of boundary conditions
of the form Eq.\re{BCcaseE} is equivalent to a set of boundary conditions
expressing $\Phi_E^u$ in terms of $\Phi_E^d$.
There are 16 real parameters for complex
fields and 10 parameters for real fields.

\item{Case F}

The following case is analogous to the preceeding one
\begin{equation}
\Phi_F^u=\left(
     \matrix{
           \phi_0\cr
           \partial\phi_0\cr
            \phi_r\cr
           \partial\phi_r\cr
           }
\right)
\label{vectorphiFu}
\end{equation}
and 
\begin{equation}
\Phi_F^d=\left(
     \matrix{
           \phi_l\cr
           \partial\phi_l\cr
            \phi_R\cr
           \partial\phi_R\cr
           }
\right)
\label{vectorphiFd}
\end{equation}
with the permutation
\begin{equation}
P_F=\left(
     \matrix{
             1& 0& 0& 0& 0& 0& 0& 0\cr
             0& 1& 0& 0& 0& 0& 0& 0\cr
             0& 0& 0& 0& 1& 0& 0& 0\cr
             0& 0& 0& 0& 0& 1& 0& 0\cr
             0& 0& 1& 0& 0& 0& 0& 0\cr
             0& 0& 0& 1& 0& 0& 0& 0\cr
             0& 0& 0& 0& 0& 0& 1& 0\cr
             0& 0& 0& 0& 0& 0& 0& 1\cr       
           }
\right)\ .
\label{PF}
\end{equation}
The boundary conditions \re{BCV} becomes
\begin{equation}
\Phi_F^d=V_F^{[4]}\Phi_F^u\ .
\label{BCcaseF}
\end{equation}
The equations are close to those of the Case C with $S_{C}^{[4]}$ replaced by
\begin{equation}
S_{F}^{[4]}=\left(
     \matrix{
           i\sigma_2&0^{[2]}\cr
           0^{[2]}&i\sigma_2
           }
\right)=1^{[4]}\otimes (i\sigma_2)\ .
\label{matrixSF4}
\end{equation}

\item{Case G}

Let us also give the results when one field boundary value and
three derivatives boundary
values are dependent variables. This is another type of boundary conditions
\begin{equation}
\Phi_G^u=\left(
     \matrix{
           \phi_0\cr
           \partial\phi_0\cr
            \phi_r\cr
           \phi_l\cr
           }
\right)
\label{vectorphiGu}
\end{equation}
and 
\begin{equation}
\Phi_G^d=\left(
     \matrix{
           \partial\phi_l\cr
           \partial\phi_l\cr
            \phi_R\cr
           \partial\phi_R\cr
           }
\right)
\label{vectorphiGd}
\end{equation}
from the permutation
\begin{equation}
P_G=\left(
     \matrix{
             1& 0& 0& 0& 0& 0& 0& 0\cr
             0& 1& 0& 0& 0& 0& 0& 0\cr
             0& 0& 1& 0& 0& 0& 0& 0\cr
             0& 0& 0& 0& 1& 0& 0& 0\cr
             0& 0& 0& 1& 0& 0& 0& 0\cr
             0& 0& 0& 0& 0& 1& 0& 0\cr
             0& 0& 0& 0& 0& 0& 1& 0\cr
             0& 0& 0& 0& 0& 0& 0& 1\cr       
           }
\right)\ .
\label{PG}
\end{equation}
The boundary conditions \re{BCV} become
\begin{equation}
\Phi_G^d=V_G^{[4]}\Phi_G^u\ .
\label{BCcaseG}
\end{equation}

The matrix $S^{[8]}_G$ is
\bea
S_{G1}=\left(
     \matrix{
           i\sigma_2 & 0^{[2]}\cr
           0^{[2]}         &0^{[2]}
           }
\right)=
&\quad,\quad&
S_{G2}=\left(
     \matrix{
           0^{[2]} & 0^{[2]}\cr
           -\sigma_3&0^{[2]}
           }
\right)
     \nonumber\\
S_{G3}=\left(
     \matrix{
           0^{[2]}&\sigma_3 \cr
           0^{[2]}         &0^{[2]}
           }
\right)
&\quad,\quad&
S_{G4}=\left(
     \matrix{
           0^{[2]} & 0^{[2]}\cr
           0^{[2]}&-i\sigma_2
           }
\right)\ .
\label{SG}
\eea
Introducing the form
\begin{equation}
V_{G}^{[4]}=\left(
     \matrix{
           V_{G1}^{[2]}&V_{G2}^{[2]}\cr
           V_{G3}^{[2]}&V_{G4}^{[2]}
           }
\right)
\label{matrixVG}
\end{equation}
in \re{SPBC}, one obtains the following restrictions on $V_{Gj}^{[2]}$
\bea
(i\sigma_2)&=&V_{G3}^{[2]+}\,(i\sigma_2)\,V_{G3}^{[2]}
     \nonumber\\
\left(\sigma_3\,V_{G2}^{[2]}\right)^+-\sigma_3\,V_{G2}^{[2]}  &=& V_{G4}^{[2]+}\,(i\sigma_2)\, V_{G4}^{[2]} 
     \nonumber\\
V_{G1}^{[2]}&=&-\sigma_3 V_{G4}^{[2]+}\,(i\sigma_2)\, V_{G3}^{[2]}\ . 
\label{conV4G}
\eea
Hence, $V_{G4}^{[2]}$ is arbitrary (8 real parameters)
and $V_{G3}^{[2]}$ is a $2 \times 2$ complex-symplectic
matrix (4 real parameters). The anti-hermitian part of
$\sigma_3V_{G2}^{[2]}$ (leaving 4 real parameters
for its hermitian part) and $V_{G1}^{[2]}$ 
are known in terms of $V_{G3}^{[2]}$ and $V_{G4}^{[2]}$. Altogether there are 16 real parameters. The
matrix $V^{[4]}_G$ is not always invertible.

\end{description}

%\newpage

\section{Reversal of the strip {\label{reversal}}}

By the transformation
\begin{equation}
s'=2\pi R - s\ ,
\label{twopiRtozero}
\end{equation}
the strip is mapped onto itself in the reversed direction with $s=\pi R$ as a fixed point. 
The end points are interchanged 
$0\leftrightarrow 2\pi R$. For a singular point at $s_1$, 
we write
\begin{equation}
\frac{s_1'}{2\pi R}=1-\frac{s_1}{2\pi R}\ .
\label{s1prime}
\end{equation}
 
The model for a singularity at $s_1$ 
is not related in a completely straightforward way to the model for 
the transformed position
$s'_1$.
It must take into account a rescaling 
of the metric which is needed to put it in our canonical form. 
Consider the metric \re{onesingular} for a single singularity at $s_1$
\bea
&{\rm{for\ }} 0\leq s\leq s_1\quad& dS^2
     =e^{-2\epsilon k s} dx_{\mu}dx^{\mu}-ds^2
     \nonumber\\
&{\rm{for\ }} s_1\leq s\leq 2\pi R\quad& dS^2
     =e^{2\epsilon k(s-2s_1)} dx_{\mu}dx^{\mu}-ds^2 \ .
\label{onesingular1}
\eea
Denote by $X$ the quantity
\begin{equation}
X=e^{2\epsilon k(s_1-\pi R)}
\label{defX}
\end{equation}
to define the rescaling ($kR=\widetilde{k}\widetilde{R}$)
\bea
\widetilde{R}&=&XR
   \nonumber\\
\widetilde{k}&=&\frac{k}{X}    
\label{defXRk}
\eea
and the change of variables
\begin{equation}
\widetilde{s}=X(2\pi R - s)\ .
\label{stilde}
\end{equation}
Consequently
\begin{equation}
\frac{\widetilde{s}_1}{2\pi \widetilde{R}}=1-\frac{s_1}{2\pi R}
\label{s1tilde2}
\end{equation}
which is analogous to \re{s1prime} but takes into account the rescaling in $R$.

The metric, rescaled in such a way that
$\widetilde{g}_{\mu\nu}=\eta_{\mu\nu}$ for $\widetilde{s}=0$ and 
$\widetilde g_{ss}=-1$ as in \re{onesingular1}, then becomes
\bea
&{\rm{for\ }} 0\leq \widetilde{s}\leq \widetilde{s}_1\quad
       & d\widetilde{S}^2=X^2 dS^2
       =e^{-2\epsilon \widetilde{k} \widetilde{s}} dx_{\mu}dx^{\mu}
       -d\widetilde{s}^2
     \nonumber\\
&{\rm{for\ }} \widetilde{s}_1\leq \widetilde{s}\leq 2\pi \widetilde{R}\quad
       & d\widetilde{S}^2=X^2 dS^2
       =e^{2\epsilon \widetilde{k}(\widetilde{s}-2\widetilde{s}_1)} dx_{\mu}dx^{\mu}
       -d\widetilde{s}^2 
\label{onesingular2}
\eea
providing the same canonical form in the tilde \re{onesingular2} 
and untilde  \re{onesingular1} 
variables.

It follows that, mutatis mutandis, the Kaluza-Klein eigenvalue 
mass towers are identical in the two cases.
Remark moreover that if $s_1=\pi R$ allowing 
in particular the closing of the strip into a circle, 
$X$ becomes exactly equal to one, the value of $s_1'$ \re{s1prime} 
become identical to the value 
of $\widetilde{s}_1$ and we are lead to
the orbifold $Z_2$ symmetry \re{twopiRtozero} of the metric  
as in the Randall Sundrum scenario.

%\newpage

\section{Scaling. Discussion of the choice $\overline{k}=1$ {\label{kvalue}}}

The mass eigenvalue equations are covariant under the rescaling 
\begin{equation}
\overline{p}\rightarrow \lambda^d \overline{p}
\label{rescaling1}
\end{equation}
of the reduced parameter
\begin{equation} 
\left\{\overline{p}\right\}\equiv\left\{\overline{k},\overline{R},\overline{s_1},\overline{\alpha_1},
\overline{\alpha_2},\overline{\alpha_3},\overline{\alpha_4},\overline{\rho_1},\overline{\rho_2},
\overline{\kappa},\overline{\zeta}\right\}\equiv\left\{\overline{k},\overline{R},\dots\right\}
\label{rescaling0}
\end{equation}
where $d$ is the energy dimension of the original parameter \re{barpara}
and $\lambda$ an arbitrary non zero real factor. Indeed, the mass eigenvalues satisfy the equation
\begin{equation}
\lambda\,m_n\left(\left\{{\overline{p}}\right\}\right)
=m_n \left(\left\{\lambda^{d}\,\overline{p}\right\}\right)\ .
\label{rescaling3}
\end{equation}
This allows one to determine the mass eigenvalues for a given 
$\overline{k}$ from the eigenvalues corresponding to our choice $\overline{k}=1$.
Choosing a rescaling with $\lambda=1/\overline{k}$,
one gets explicitly
\begin{equation}
m_n\left( \left\{ \overline{k},\overline{R},\dots \right\}\right)
=\overline{k}\, m_n \left(\left\{1,\overline{k}\,\overline{R},\dots\right\}\right) \ .
\label{rescaling2}
\end{equation}

\newpage

\vskip 0.5 cm
\begin{table}
\caption{
Allowed boundary conditions for the strip case. With a single 
periodic or antiperiodic metric singularity at $s_1$ 
($0<s_1<2\pi R$), $T=e^{-4 \epsilon k(\pi R-s_1)}$
(see \re{Tdef1} and \re{transitioni}). 
{\label{tableBCdeux}}
}
\vspace{1 cm}
\hspace{1 cm}
\scriptsize
{
\begin{tabular}{|l|l|l|}
\hline
\multicolumn{3}{|c|}
      {Two Boundary Conditions}
        \\ \hline
    Case& Boundary Conditions&
      \\ \hline\hline
    A1            & $\phi(2\pi R)=T\left(\alpha_1\phi(0)
                     +\alpha_2\partial_s\phi(0)\right) $  &$\alpha_2\neq 0$ 
                       \\
                    & $\partial_s\phi(2\pi R)=T
                    \left(\frac{\alpha_1\alpha_4-1}
                    {\alpha_2}\phi(0)
                    +\alpha_4\partial_s\phi(0)\right) $   &
       \\ \hline
    A2            & $\phi(2\pi R)=T\alpha_1\phi(0)$ &$\alpha_1\neq 0$  
                       \\
                    & $\partial_s\phi(2\pi R)=T
                    \left(\alpha_3\phi(0)
                    +\frac{1}{\alpha_1}\partial_s\phi(0)\right) $&
       \\ \hline
    A3              & $\partial_s\phi(0)=\rho_1\phi(0) $ &
                       \\
                    & $\partial_s\phi(2\pi R)=\rho_2\phi(2\pi R) $ &
       \\ \hline
    A4              & $\phi(0)=0 $      &
                       \\
                    & $\partial_s\phi(2\pi R)=\kappa\phi(2\pi R) $   &
       \\ \hline
    A5              & $\phi(2\pi R)=0 $       &
                       \\
                    & $\partial_s\phi(0)=\zeta\phi(0) $   &
       \\ \hline
    A6              & $\phi(0)=0 $        &
                       \\
                    & $\phi(2\pi R)=0 $   &
      \\ \hline
\end{tabular}
   }
\end{table}

\vskip 0.5 cm
\begin{table}
\caption{Zero mass conditions 
between the boundary condition parameters 
and $y_1$ for a model with
a single periodic or antiperiodic 
metric singularity ($\delta_1=\pm 1$) at $s_1=y_1 \pi R$ ($0\leq y_1\leq 2$);
$E=e^{2 \epsilon k \pi R}$
and $F = E^{2(1-y_1)}-2E^{2}+E^{-2(1-y_1)}$.
{\label{tablemasszero}}
}
\vspace{1 cm}
\hspace{1 cm}
\scriptsize
{
\begin{tabular}{|l|l|}
\hline
\multicolumn{2}{|c|}
      {Zero mass conditions}
        \\ \hline
    Case& Parameter conditions
      \\ \hline\hline
        &\\
    A1     &          $
           \left(\alpha_1\alpha_4-1\right) F
            +4\epsilon k \alpha_2\left(\alpha_1 E^{-2(1-y_1)}+\alpha_4 E^{2(1-y_1)}-2\delta_1\right)
           =0          $
                       \\
           &          
              \quad \quad one solution
                       \\
                     &  \\
                    \cline{2-2}
                    &\\
           &  $\alpha_1= \delta_1 E^{2(1-y_1)}$       
                      \\ 
           &  $\alpha_4 = \delta_1 E^{-2(1-y_1)} $      
                      \\       
           & $\alpha_2 =-
                \frac{\epsilon\delta_1 F}
                 {4k  }      $        
                      \\
                      &
                             \quad \quad two independent solutions
                      \\
                      & 
          \\ \hline\hline
                     &  \\ 
    A2          &   
           $ \alpha_1 \alpha_3 F
           +4\epsilon k\left(\alpha_1 \delta_1 E^{-(1-y_1)}-E^{(1-y_1)}\right)^2
                             $ 
                       \\
                &  
                     \quad \quad one solution  \\
                              &\\ 
                    \cline{2-2}
                &    \\ 
       &  $\alpha_1=\delta_1 E^{2(1-y_1)}$  
                       \\       
                      &   $\alpha_3=0$ 
                      \\
                   &  $ F= 0  $      
                     \\
                     & 
                   \quad \quad two independent solutions
                   \\ 
                 &
       \\ \hline\hline
      &\\    
    A3       &       $
             \rho_1 \rho_2 F+4 \epsilon k\left(\rho_1 E^{-2(1-y_1)}-\rho_2 E^{2(1-y_1)}\right)                                          =0$       
                    \\
                     & 
                   \quad \quad one solution\\ 
                                            &  \\ 
             \hline\hline
       &\\
    A4       &  $
               \kappa F +4\epsilon k E^{-2(1-y_1)} = 0
                      $
                       \\
                     & 
                   \quad \quad one solution\\ 
                                            &  \\ 
              \hline\hline
       &\\
    A5      &         $
              \zeta F -4 \epsilon k E^{2(1-y_1)} =0         
                      $
                       \\
                     & 
                   \quad \quad one solution\\ 
            &        
       \\ \hline\hline
       &\\
    A6     &          $
                F=0
                      $
                       \\
                     & 
                   \quad \quad one solution\\ 
           &         
      \\ \hline\hline
\end{tabular}
   }
\end{table}

\begin{table}
\caption{Low lying Kaluza-Klein masses 
in the Case A2 ($\alpha_1=1,\ \alpha_3=0$), for
$\overline{k}=1$ and
$k R=6.3$, as a function of the position $s_1$ of a 
periodic singularity. Masses are in TeV.
{\label{tableKK63}}
}
\vspace{1 cm}
\hspace{1 cm}
\scriptsize
{
\begin{tabular}{|c|c|c|l|l|l|l|l|}
\hline
 $kR$ & $y_1=s_1/(\pi R)$&$m_1$&$m_2$&$m_3$&$m_4$&$m_5$&$m_6$
     \\ \hline
 6.3  &  2                  &0.30 &0.55&0.80	&1.05&1.29&1.54   
     \\ \hline
 6.3  &  1.995              &0.29 &0.54&0.79	&1.04&1.29&1.54   
     \\ \hline
 6.3  &  1.99               &0.28 &0.53&0.80	&1.07&1.33&1.59   
     \\ \hline
 6.3  &  1.97               &0.21 &0.65&1.00	&1.23&1.60&1.93   
     \\ \hline
 6.3  &  1.95               &0.18 &0.90&1.30	&1.70&2.14&2.55
      \\ \hline
 6.3  &  1.90               &0.16 &2.23&3.02	&4.13&5.00&6.03   
      \\ \hline
 6.3  &  1.85               &0.16 &5.90&7.90	&10.80&12.98&15.70   
      \\ \hline
 6.3  &  1.80               &0.16 &15.8&21.2&28.9&34.7&41.9   
      \\ \hline
 6.3  &  1.70               &0.16 &114&153	&204 &251 &302   
     \\ \hline
 6.3  &  1.60               &0.16 &826&1107	&1511&1813&2200
      \\ \hline
 6.3  &  1.50               &0.16 &6000&8000	&10900&13100&15900   
       \\ \hline
\end{tabular}
   }
\end{table}

\begin{table}
\caption{Low lying Kaluza-Klein masses 
in the Case A2 ($\alpha_1= 1,\ \alpha_3=0$), for 
$\overline{k}=1$ and
a periodic singularity at
$s_1=1.5\pi R$, 
as a function of  $kR$. Masses are in TeV.
{\label{tableKK15}}
}
\vspace{1 cm}
\hspace{1 cm}
\scriptsize
{
\begin{tabular}{|c|c|c|c|l|l|l|l|l|}
\hline
 $kR$ & $y_1=s_1/(\pi R)$&$m_1$&$m_2$&$m_3$&$m_4$&$m_5$&$m_6$
     \\ \hline
$\bigl.\bigr.$ 6.3  &  1.5            &16\ $10^{-2}$ &6000&8000	&10900&13100&15900   
     \\ \hline
$\bigl.\bigr.$ 6.4  &  1.5            &8.4\ $10^{-2}$  &3700&5000	&6800&8200&9900   
     \\ \hline
$\bigl.\bigr.$ 6.6  &  1.5            &2.4\ $10^{-2}$ &1500&2000	&2700&3200&3900   
     \\ \hline
$\bigl.\bigr.$ 6.8  &  1.5            &6.2\ $10^{-3}$&570  &760 	&1000&1200&1500   
     \\ \hline
$\bigl.\bigr.$ 7.0  &  1.5            &1.9\ $10^{-3}$&220  &300	  &400  &490 &590
      \\ \hline
$\bigl.\bigr.$ 7.5  &  1.5            &8.3\ $10^{-5}$&21   &28	  &38   &46  &56   
      \\ \hline
$\bigl.\bigr.$ 7.9  &  1.5            &6.8\ $10^{-6}$&3.2  &4.3	 &5.8   &6.0 &8.4   
      \\ \hline
$\bigl.\bigr.$ 8.3  &  1.5            &5.4\ $10^{-7}$&0.48 &0.64 &0.88  &1.06&1.28   
       \\ \hline
$\bigl.\bigr.$ 8.4  &  1.5            &2.9\ $10^{-7}$&0.30 &0.40 &0.55  &0.66&0.80   
       \\ \hline
\end{tabular}
   }
\end{table}

\begin{table}
\caption{Low lying Kaluza-Klein masses in the Case A2 ($\alpha_1=1,\ \alpha_3=0$), 
with $\overline{k}=1$ and  $kR=12.6/y_1$, 
as a function of the position $s_1=y_1\pi R$ of a periodic singularity. 
Masses are in TeV.
{\label{tableKK126}}
}
\vspace{1 cm}
\hspace{0.5 cm}
\scriptsize
{
\begin{tabular}{|c|c|c|c|l|l|l|l|l|l|}
\hline
 $kR=12.6/y_1$ & $y_1=s_1/(\pi R)$&$m_1$&$m_2$&$m_3$&$m_4$&$m_5$&$m_6$&$m_7$
     \\ \hline
$\bigl.\bigr.$ 6.3  &  2              &0.30         &0.55&0.80	&1.05&1.29&1.54&1.79   
     \\ \hline
$\bigl.\bigr.$ 6.332  &  1.99         &0.23         &0.43&0.65	&0.87&1.09&1.3&1.51   
     \\ \hline
$\bigl.\bigr.$ 6.46  &  1.95          &0.064        &0.33&0.48	&0.63&0.79&0.93&1.09   
     \\ \hline
$\bigl.\bigr.$ 6.63  &  1.90          &1.98\ $10^{-2}$&0.31  &0.42 	&0.57&0.69&0.84&0.96   
     \\ \hline
$\bigl.\bigr.$ 7.0  &  1.80           &1.93\ $10^{-3}$&0.301&0.404&0.551&0.662&0.800&0.914
      \\ \hline
$\bigl.\bigr.$ 8.4  &  1.50           &2.92\ $10^{-7}$&0.301&0.403&0.551&0.661&0.799&0.913  
      \\ \hline
$\bigl.\bigr.$ 11.45  &  1.10         &1.5\ $10^{-15}$&0.301&0.403&0.551&0.661&0.799&0.913  
      \\ \hline
$\bigl.\bigr.$ 12.594  &  1.0005      &5.8\ $10^{-20}$&0.301&0.403&0.551&0.661&0.799&0.913   
       \\ \hline
$\bigl.\bigr.$ 12.6  &  1.0           &0              &0.301&0.403&0.551&0.661&0.799&0.913   
       \\ \hline
$\bigl.\bigr.$ 12.606  &  0.9995     &5.5\ $10^{-20}$&0.301&0.403&0.551&0.661&0.799&0.913   
       \\ \hline
$\bigl.\bigr.$ 14.0  &  0.9          &1.01\ $10^{-18}$&0.301&0.403&0.551&0.661&0.799&0.913  
       \\ \hline
$\bigl.\bigr.$ 15.75  &  0.5         &1.01\ $10^{-18}$&0.301&0.403&0.551&0.661&0.799&0.913   
       \\ \hline
$\bigl.\bigr.$ 126  &  0.1           &1.01\ $10^{-18}$&0.301&0.403&0.551&0.661&0.799&0.913
       \\ \hline
$\bigl.\bigr.$ 252  &  0.05          &1.01\ $10^{-18}$&0.301&0.403&0.551&0.661&0.799&0.913   
       \\ \hline
\end{tabular}
   }
\end{table}

\begin{table}
\caption{Low lying Kaluza-Klein masses in the Case A2 ($\alpha_1=1$)
with a periodic singularity at $s_1=y_1\pi R,\ y_1=1.95$, 
for  $\overline{k}=1$ and $kR=12.6/y_1$,
as a function of ${\overline{\alpha_3}}$.
Masses are in TeV.
{\label{tableKK195}}
}
\vspace{1 cm}
\hspace{1 cm}
\scriptsize
{
\begin{tabular}{|c|c|c|c|l|l|l|l|l|}
\hline
 ${\overline{\alpha_3}}$ & $h$&$m_1$&$m_2$&$m_3$&$m_4$&$m_5$&$m_6$&$m_7$
     \\ \hline
$\bigl.\bigr.$ -100  &         &0.309         &0.425&0.587	&0.729&0.874&1.031&1.170   
     \\ \hline
$\bigl.\bigr.$ -10   &         &0.307         &0.419&0.579	&0.713&0.860&1.009&1.148   
     \\ \hline
$\bigl.\bigr.$ -1    &         &0.266         &0.362&0.511	&0.643&0.806&0.944&1.099   
     \\ \hline
$\bigl.\bigr.$ -0.2  &         &0.158         &0.338&0.486 	&0.630&0.795&0.934&1.092   
     \\ \hline
$\bigl.\bigr.$ -0.1  &         &0.123         &0.336&0.483  &0.629&0.794&0.933&1.091
      \\ \hline
$\bigl.\bigr.$ 0.01  &         &0.054         &0.334&0.480  &0.627&0.792&0.931&1.090  
      \\ \hline
$\bigl.\bigr.$ 0.03  &         &0.024         &0.333&0.479  &0.627&0.792&0.931&1.090  
      \\ \hline
$\bigl.\bigr.$ 0.0347&         &0.0035        &0.333&0.479  &0.627&0.792&0.931&1.091   
       \\ \hline
$\bigl.\bigr.$ 
$\overline{\alpha}_3^{[0]}$=0.034815127...&    &0             &0.333&0.479  &0.627&0.792&0.931&1.090   
       \\ \hline
$\bigl.\bigr.$ 0.034803&  0.000421&           &0.333&0.479  &0.627&0.792&0.931&1.090   
       \\ \hline
$\bigl.\bigr.$ 0.03481 &  0.0010  &           &0.333&0.479  &0.627&0.792&0.931&1.090  
       \\ \hline
$\bigl.\bigr.$ 0.0349  &  0.0034  &           &0.333&0.479  &0.627&0.792&0.931&1.090   
       \\ \hline
$\bigl.\bigr.$ 0.04  &  0.0249    &           &0.332&0.478  &0.627&0.791&0.931&1.090
       \\ \hline
$\bigl.\bigr.$ 0.05  &  0.0420    &           &0.332&0.477  &0.626&0.791&0.930&1.090   
       \\ \hline
$\bigl.\bigr.$ 0.1  &  0.0894     &           &0.332&0.477  &0.626&0.791&0.930&1.090   
       \\ \hline
$\bigl.\bigr.$ 0.5  &  0.259      &           &0.326&0.468  &0.621&0.786&0.926&1.086   
       \\ \hline
$\bigl.\bigr.$ 1.0  &  0.406      &           &0.322&0.460  &0.617&0.780&0.921&1.083   
       \\ \hline
$\bigl.\bigr.$ 10  &  2.460       &           &0.311&0.432  &0.595&0.745&0.889&1.051   
       \\ \hline
$\bigl.\bigr.$ 50  &  11.16       &           &0.310&0.429  &0.592&0.739&0.883&1.043   
       \\ \hline
$\bigl.\bigr.$ 100  &  22.0       &           &0.309&0.426  &0.589&0.732&0.877&1.035   
       \\ \hline
\end{tabular}
   }
\end{table}

\begin{table}
\caption{Low lying Kaluza-Klein masses in the Case A2 
($\alpha_1=1,\overline{\alpha}_3=0$) with
a periodic singularity located at $s_1=\pi R $, for
$\overline{k}=1/100$ and  
$kR=11$ (see Sec.\re{comparison}).
The masses $m_{2n+1},\ n\geq 1$ are missing in Rizzo's tower.
Masses are in TeV.
{\label{RizzoGN}}
}
\vspace{1 cm}
\hspace{ 0.3 cm}
\scriptsize
{
\begin{tabular}{|c|l|l|l|l|l|l|l|l|l|l|}
\hline
$\bigl.\bigr.$           &$m_1$ &$m_2$   &$m_3$   &$m_4$   &$m_5$ &$m_6$&$m_7$   &$m_8$  &$m_9$&$m_{10}$ 
     \\ \hline
$\bigl.\bigr.$ Rizzo &0 &0.459&        &0.840& &1.218&      &1.595&  & 1.972  
     \\ \hline
$\bigl.\bigr.$ Us    &0 &0.459&0.615&0.840 &1.008&1.218&1.391&1.595&1.772&1.972  
     \\ \hline
\end{tabular}
   }
\end{table}

\begin{table}
\caption{Low lying Kaluza-Klein mass towers 
in the Case A5 
for $\overline{k}=1$ and $kR=12.6/y_1$,
as a function of the
position of a periodic singularity
$s_1=y_1\pi R$. The towers are independent of $\overline{\zeta}$.
In addition to these towers there is a lonely state or a tachyon
which is moving with $\overline{\zeta}$ independently of $y_1$
and given in Table \re{caseA5b}. Masses are in TeV.
{\label{caseA5a}}
}
\vspace{1 cm}
\hspace{1 cm}
\scriptsize
{
\begin{tabular}{|c|l|l|l|l|l|l|}
\hline
$\bigl.\bigr.$     $y_1$   &$m_1$  &$m_2$ &$m_3$  &$m_4$  &$m_5$  & $m_6$ 
     \\ \hline
$\bigl.\bigr.$       2     & 0.403 & 0.661 & 0.913 & 1.162 & 1.411 & 1.659
     \\ \hline			         
$\bigl.\bigr.$    1.999    & 0.396 & 0.649 & 0.896 & 1.141 & 1.385 & 1.629       
     \\ \hline			        
$\bigl.\bigr.$    1.998    & 0.390 & 0.639 & 0.881 & 1.123 & 1.363 & 1.602    
     \\ \hline			        
$\bigl.\bigr.$    1.99     & 0.360 & 0.583 & 0.794 & 0.996 & 1.198 & 1.406  
     \\ \hline			        
$\bigl.\bigr.$    1.98     & 0.339 & 0.526 & 0.692 & 0.880 & 1.075 & 1.259  
     \\ \hline			        
$\bigl.\bigr.$    1.95     & 0.309 & 0.426 & 0.588 & 0.730 & 0.875 & 1.032
      \\ \hline			        
$\bigl.\bigr.$    1.9      & 0.301 & 0.403 & 0.551 & 0.661 & 0.799 & 0.913
      \\ \hline			        
$\bigl.\bigr.$    1.5      & 0.301 & 0.403 & 0.551 & 0.661 & 0.799 & 0.913
      \\ \hline			        
$\bigl.\bigr.$    1        & 0.301 & 0.403 & 0.551 & 0.661 & 0.799 & 0.913
       \\ \hline		        
$\bigl.\bigr.$    0.1      & 0.301 & 0.403 & 0.551 & 0.661 & 0.799 & 0.913
       \\ \hline		        
$\bigl.\bigr.$    0.01     & 0.301 & 0.403 & 0.551 & 0.661 & 0.799 & 0.913
	\\ \hline	        
$\bigl.\bigr.$  $10^{-6}$  & 0.301 & 0.403 & 0.551 & 0.661 & 0.799 & 0.913
       \\ \hline
\end{tabular}					            
   }		
\end{table}

\begin{table}
\caption{Extra Kaluza-Klein lonely mass state or tachyon in the Case A5
for $\overline{k}=1$ and $kR=12.6/y_1$,
as a function of the $\overline{\zeta}$ independently of the position 
of the periodic singularity $s_1=y_1\pi R$.
This state is superimposed on the 
main Kaluza-Klein towers (Table\re{caseA5a}) which depend on $y_1$ but not on $\overline{\zeta}$.
In the table, $\overline{\zeta}^{[0]}$ is the condition for a zero mass state.
Masses are in TeV.
{\label{caseA5b}}
}
\vspace{1 cm}
\hspace{ 5 cm}
\scriptsize
{
\begin{tabular}{|c|l|l|}
\hline
$\bigl.\bigr.$ $\overline{\zeta}$& $ m$         & $h$ 
     \\
$\bigl.\bigr.$  unit $10^{-35}$  &              &
     \\ \hline
$\bigl.\bigr.$  10000            &5.456         &
     \\ \hline
$\bigl.\bigr.$  1000             &1.725         &
     \\ \hline
$\bigl.\bigr.$   700             &1.444         &
     \\ \hline
$\bigl.\bigr.$   500             &1.220         &
     \\ \hline
$\bigl.\bigr.$   300             &0.945         &
     \\ \hline
$\bigl.\bigr.$   200             &0.771         &   
     \\ \hline
$\bigl.\bigr.$   100             &0.545         &   
     \\ \hline
$\bigl.\bigr.$   50              &0.386         &   
     \\ \hline
$\bigl.\bigr.$   10              &0.172         &   
     \\ \hline
$\bigl.\bigr.$   5               &0.122         & 
     \\ \hline
$\bigl.\bigr.$   1               &0.055         &    
     \\ \hline
$\bigl.\bigr.$   0.5             &0.038         &    
     \\ \hline
$\bigl.\bigr.$   0.1             &0.017         &   
     \\ \hline
$\bigl.\bigr.$   0.001           &0.001         & 
     \\ \hline
$\bigl.\bigr.$  $\overline{\zeta}^{[0]}= 4 E^{2(1-y_1)}/F$&0     &
     \\ \hline
$\bigl.\bigr.$   -1            &              &0.0545 
     \\ \hline
$\bigl.\bigr.$   -100          &              &0.5456   
     \\ \hline
$\bigl.\bigr.$   -1000         &              &1.725 
     \\ \hline
$\bigl.\bigr.$   -10000        &              &5.456   
     \\ \hline
\end{tabular}
   }
\end{table}
\vfill\break

\begin{figure}[ht]	     
\caption{Illustration of a potential Kaluza-Klein mass spectrum
inspired by Figure 5 of Rizzo \cite{Rizzo} 
for the Case A2 with closure into a circle
($\alpha_1{=}\delta_0{=}\delta_1{=}1,\ \alpha_3{=}0,\,\ \overline{k}{=}0.01,\ kR{=}11$).
The observer is at $s_{\rm{phys}}=0$.
See Sec.\re{comparison} 
and Table \re{RizzoGN}.  
The cross section is in
arbitrary units. In the absence of knowledge 
of production and decay mechanisms 
for the Kaluza-Klein
states, the widths have been arbitrarily set to zero. 
The masses are superposed on a 
Drell-Yan type background.
Compared to the figure of Rizzo there are twice as many states in the tower.
The masses are in GeV. 
\label{Rizzofig1}	        
       }	       
\vskip 2 cm	       
\begin{center}	        
\epsfxsize=12cm		      
\epsffile{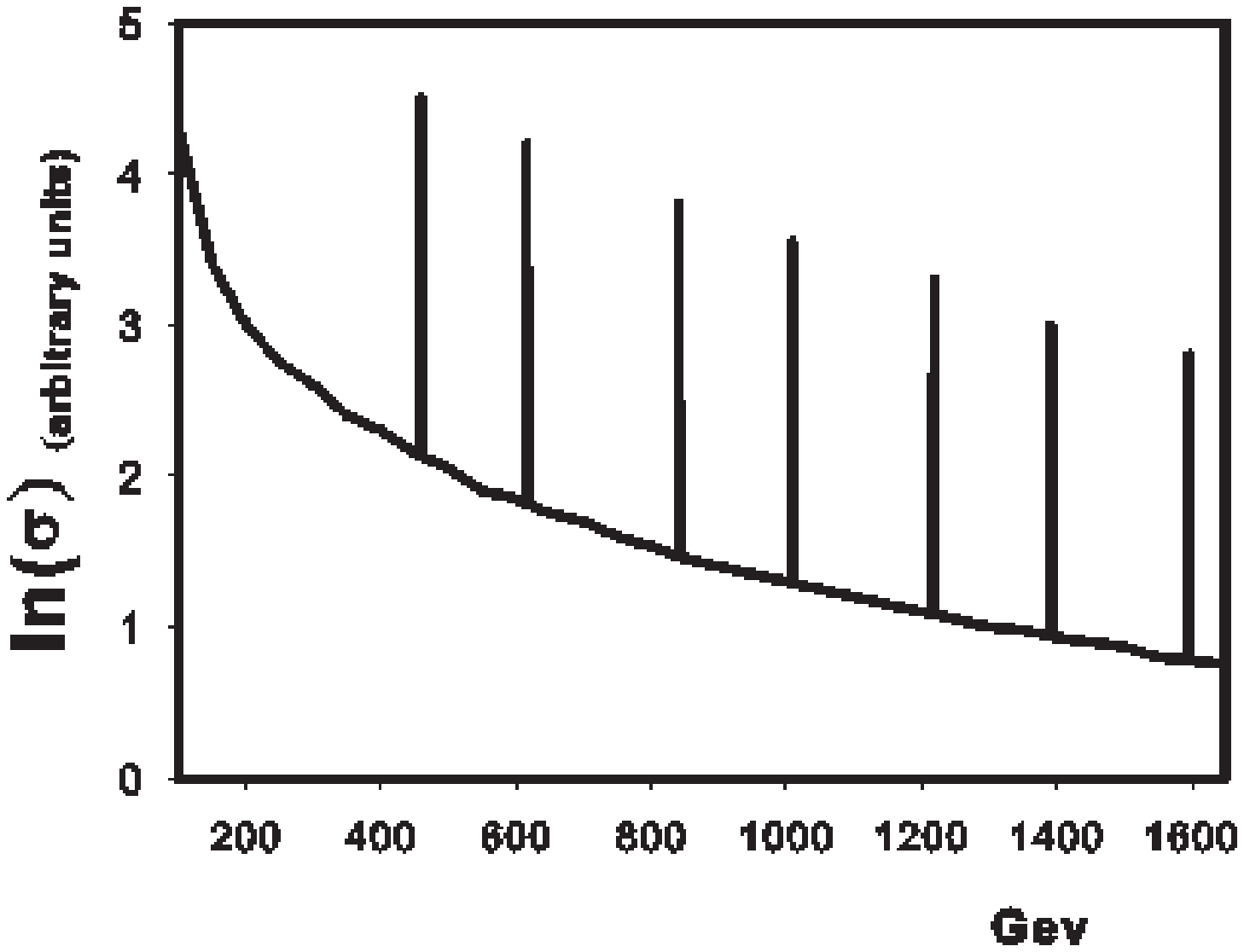}
\end{center}
\end{figure}


\newpage

\begin{thebibliography}{99}

\bibitem{GN1}
Grard,~F, Nuyts,~J.,
%{\it{Elementary Kaluza-Klein towers revisited}},
{\it Phys. Rev. D} {\bf{74}}, 124013 (2006),
hep-th/0607246

\bibitem{GN3}
Grard,~F, Nuyts,~J.,
%{\it{Kaluza-Klein towers for spinors in flat space}},
{\it Phys. Rev. D} {\bf{78}}, 024020 (2008),
hep-th/0803.1741

\bibitem{GN2}
Grard,~F, Nuyts,~J.,
%{\it{Warped Kaluza-Klein towers revisited}},
{\it Phys. Rev. D} {\bf{76}}, 124022 (2007),
hep-th/0707.4562

\bibitem{RS2}
Randall,~L., Sundrum,~R.,
%{\it{An Alternative to Compactification}},
{\it Phys. Rev. Lett.} {\bf{83}}, 4690 (1999),
hep-th/9906064


\bibitem{Ben}
Bender,~C.M., Stefan Boettcher,~S.,
%{\it{Real Spectra in Nonhermitian Hamiltonians having PT symmetry}},
{\it{Phys.Rev.Lett.}} {\bf{80}}, 5243-5246 (1998),
Physics/9712001

\bibitem{FN}
Fairlie,~D.B., Nuyts,J.,
%{\it{Fock Space Representation for Non-hermitian Hamiltonians}},
{\it{J.Phys.}} {\bf{A38}}, 3611-3624 (2005),
hep-th/0412148


\bibitem{KK}
Kaluza,~T.,
%{\it{On the problem of unity in physics}},
{\it{Sitzungsber. Preuss. Akad. Wiss. Berlin. (Math. Phys.)}}, K1, 966-972 (1921).
Klein,~O.,
%{\it{Quantum theory and five-dimensional theory of relativity}},
{\it{Z. Phys.}} {\bf{37}}, 895-906 (1926).

\bibitem{EPR}
Einstein,~A., Podolsky,~B. and N. Rosen,~N.,
%{\it{Can Quantum-Mechanical Description of Physical Reality Be Considered Complete?}},
{\it{Phys. Rev.}} {\bf{47}}, 777 (1935).

\bibitem{Rizzo}
Rizzo, Thomas G.
%{\it{Pedagogical introduction to extra dimensions}},
Proceedings of 32nd SLAC Summer Institute on Particle Physics: Natures Greatest Puzzles, Menlo Park, California, 
SLAC-PUB-10753, SSI-2004-L013,  
hep-ph/0409309 

\bibitem{RS1}
Randall,~L., Sundrum,~R.,
%{\it{A Large Mass Hierarchy from a Small Extra Dimension}},
{\it Phys. Rev. Lett.} {\bf{83}}, 3370 (1999),
hep-ph/9905221,

\end{thebibliography}
\end{document}